\documentclass[a4paper]{article}

\usepackage{doc}
\usepackage{makeidx}
\usepackage{float}
\usepackage[tight,footnotesize]{subfigure}
\usepackage[dvips]{graphicx}
\usepackage{graphicx}
\usepackage{amsmath}
\usepackage{breqn}
\usepackage{multirow}
\usepackage{algorithm}
\usepackage{algpseudocode}
\usepackage{adjustbox}
\usepackage{url}

\usepackage{listings}
\usepackage{color}
\usepackage{authblk}
 
\definecolor{codeblack}{rgb}{0,0.6,0}
\definecolor{codegray}{rgb}{0.5,0.5,0.5}
\definecolor{codepurple}{rgb}{0.58,0,0.82}
\definecolor{backcolour}{rgb}{0.95,0.95,0.92}
\definecolor{white}{rgb}{1.0, 1.0, 1.0}
\definecolor{black}{rgb}{0.0, 0.0, 0.0}

\lstdefinestyle{mystyle}{
    backgroundcolor=\color{white},
    commentstyle=\color{codegray},
    keywordstyle=\color{black},
    numberstyle=\tiny\color{codegray},
    stringstyle=\color{codepurple},
    basicstyle=\footnotesize,
    breakatwhitespace=false,
    breaklines=true,
    captionpos=b,
    keepspaces=true,
    numbers=left,
    numbersep=5pt,
    showspaces=false,
    showstringspaces=false,
    showtabs=false,
    tabsize=2,
	mathescape=true    
}
 
\newcommand{\keywords}[1]{ \medskip \noindent \textbf{Keywords:} #1 }

\newcommand{\CommentLine}[1]{
    \State  \textcolor{black}{\texttt{/* #1 */ } }
}
\newcommand{\CommentSingleLine}[1]{
      \hfill \textcolor{black}{\texttt{// #1 }} 
}

\title {Soft Errors Detection and
Automatic Recovery based on Replication combined with different Levels of Checkpointing}

\author[1]{Diego Montezanti}
\author[1]{Enzo Rucci}
\author[1]{Armando De Giusti}
\author[1]{Marcelo Naiouf}
\author[2]{Dolores Rexachs}
\author[2]{Emilio Luque}

\affil[1]{III-LIDI - Instituto de Investigaci\'on en Inform\'atica LIDI \\ Facultad de Inform\'atica,
Universidad Nacional de La Plata\\
La Plata, Buenos Aires, Argentina\\
 \authorcr
\{dmontezanti,erucci,degiusti,mnaiouf\}@lidi.info.unlp.edu.ar} 
\affil[2]{CAOS - Computer Architecture and Operating Systems\\
Universidad Aut\'onoma de Barcelona, Barcelona, Spain\\
\authorcr
\{dolores.rexachs,emilio.luque\}@uab.es}

\date{{June 8, 2020}}

\begin{document}

\maketitle 

\begin{center}
\texttt{This is the accepted version of the manuscript that was sent to review to \textit{Future Generation Computer Systems} (ISSN 0167-739X). This manuscript was finally accepted for publication on July 3rd, 2020 and its final published version is available online at \url{https://doi.org/10.1016/j.future.2020.07.003}}.
\end{center}

\begin{center}
\texttt{\textregistered 2020. This manuscript version is made available under the CC-BY-NC-ND 4.0 license \url{http://creativecommons.org/licenses/by-nc-nd/4.0/}}.
\end{center}

\clearpage

\begin{abstract}

Handling faults is a growing concern in HPC. {\color{black}In future exascale systems, it is projected that silent undetected errors will occur several times a day, increasing the occurrence of corrupted results}. In this article, we propose SEDAR, {\color{black}which is} a methodology that improves system reliability against transient faults when running parallel message-passing applications. {\color{black}Our approach}, based on process replication for detection, combined with different levels of checkpointing for automatic recovery, has the goal of helping users of scientific applications to {\color{black}obtain executions} with correct results. SEDAR is structured in three levels: (1) only detection and safe-stop with notification; (2) recovery based on multiple system-level checkpoints; and (3) recovery based on a single valid user-level checkpoint. {\color{black}As e}ach of these variants supplies a particular coverage but {\color{black}involves} limitations and implementation costs, {\color{black}SEDAR can be adapted to the needs of the} system. {\color{black}In this work}, a description of the methodology {\color{black}is presented and the temporal behavior of employing each SEDAR strategy is mathematically described, both in the absence and presence of faults. A model that considers all the fault scenarios on a test application is introduced to show the validity of the detection and recovery mechanisms. An overhead evaluation of each variant is performed with applications involving different communication patterns; this is also used to extract guidelines about when it is beneficial to employ each SEDAR protection level}. As a result, we show its efficacy and viability to tolerate transient faults in target HPC environments.

\end{abstract}


\section{Introduction}
\label{sect:introduction}
In the area of High-Performance Computing (HPC), parallel systems continue increasing the number of components to improve their performance and, as a consequence, ensuring their reliability has become a critical issue. Nowadays, fault rates involve just a few hours on modern platforms \cite{martsin2015} but it is forecasted that large parallel applications will have to manage fault rates of barely some minutes in exascale supercomputers \cite{benoit2019combining}. In that sense, these applications require some help to progress efficiently.

Currently, there are different types of transient errors affecting parallel programs; Silent Data Corruption (SDC) is the most dangerous of {\color{black}these} as several recent reports have stated~\cite{elliott2014}. When SDC occurs, the application seems to run correctly but, at the end, the results are {\color{black}incorrect}. Science is one of the areas strongly affected by SDC, since historically {\color{black}it} has relied on large-scale simulations. Therefore, the treatment of silent errors is one of the greatest challenges in current and future resilience.

Without a fault tolerance mechanism, a whole application could {\color{black}misbehave} due to a failure affecting just one task. {\color{black}Even worse}, the program could output invalid results that, in the best-case scenario, will be noticed {\color{black}when the execution is concluded}{\color{black}; as a consequence, silent errors require a detection mechanism \cite{benoit2018coping}}. In addition, a single SDC can cause deep effects, propagating them across all processes that communicate in message-passing applications \cite{fiala2012}. One way to tolerate transient faults is to rely on hardware redundancy (registers and processor arithmetical logic units){\color{black}. N}evertheless, this approach is highly expensive and difficult to implement \cite{shye2009}. Given the high cost of re-running the application from the start if a fault is detected, specific software strategies are required to reach a suitable cost-benefit trade-off. 

{\color{black}As opposed to the silent errors, the} fail-stop failures cause a process to crash{\color{black},} making their detection almost immediate. A common, well-studied technique to reduce their impact consist{\color{black}s} of Checkpoint-based Rollback recovery (C/R) \cite{cappello2014}. When coordinated checkpointing is used, the entire state of an application is saved in a periodic manner. So, if a failure takes place, all the processes can restart from their saved checkpoints. {\color{black} Instead, if uncoordinated checkpointing is used, only the state of the process being checkpointed is dumped.} Unfortunately, C/R can be time-consuming and the overhead increases as the number of cores grow. Nonetheless, {\color{black}despite being effective {\color{black}when dealing} with fail-stop errors, C/R shows weakness when facing silent errors. Since a stored checkpoint could contain undetected corruption, C/R {\color{black}cannot} guarantee a correct recovery}. This situation {\color{black}becomes} aggravated in the case of strongly coupled computation, since an error in one node could propagate to the others in microseconds \cite{lu2013}. 

Performing redundant software execution is a common way to provide resilience. Following the state machine replication approach, a process is duplicated and both copies proceed with the same execution sequence. As a result, for deterministic applications, they produce the same output for the same input \cite{mushtaq2013}{\color{black}; these two outputs can be compared to provide error detection, despite not being enough for recovery}. In the HPC context, using multicore architectures represents a viable solution for detecting SDCs {\color{black}as a result of} their {\color{black}intrinsic} natural redundancy \cite{martsin2015}.

Considering these circumstances of non-reliable results and their expensive verification, this paper presents SEDAR, which is a methodology designed to provide transient fault tolerance for scientific message-passing parallel applications that execute in multicore clusters. SEDAR seeks to help programmers and users of parallel scientific applications to accomplish reliability in their executions. {\color{black}It works as a static library that is compiled with the application{\color{black}. Even though this changes} the model of execution, {\color{black}it is still} almost transparent to the algorithm, as {\color{black}opposed} to specific detectors that {\color{black}force modifying} it and do not cover all faults \cite{benoit2018coping,bosilca2009algorithm}}. Following this methodology, each process of the parallel application gets replicated and both copies {\color{black}execute} on different cores of the same socket, taking advantage of the multicore's {\color{black} intrinsic} hardware redundancy. {\color{black} SEDAR can detect and recover from all transient faults that cause SDC and TOE (Time Out Errors)}. Three different ways are provided by SEDAR so it can achieve {\color{black}full} silent error coverage: (1) only detection with notification; (2) recovery based on multiple system-level checkpoints; and (3) recovery utilizing a single safe application-level checkpoint. Each of these alternatives has particular features and provides a different cost-performance trade-off.


A preliminary, more conceptual version of this work is available in \cite{montezanti2017methodology}. This article fully describes and validates the methodology, extending the insights already offered in the previous {\color{black}version}, with the following new contributions:

\begin{itemize}

\item {\color{black}The introduction of an analytical model to verify the efficacy both of the detection strategy and of the recovery mechanism based on multiple, system-level coordinated checkpoints}. The model is based on the predictability of the data affected in a well-known test application (considering its computation and communication stages), so the consequences of each fault occurrence can be anticipated.
\item The design of a complete workfault {\color{black}(i.e. a set of representative fault cases that emulate real faults experienced by the system, which we use as workload for testing purposes) that permits a grouping of all possible faulty situations in \textit{scenarios}}. This set includes the effects of the faults, their latency of detection and recovery point. 
\item The implementation of the multiple-checkpoint-based recovery algorithm, using the DMTCP library. {\color{black}To verify its operation, SEDAR has been incorporated into the test application. The empirical validation has been achieved through controlled fault injection}.
\item {\color{black} The evaluation of the overhead of each alternative SEDAR strategy. For this purpose, we} have attached SEDAR to three parallel benchmarks with different communication patterns and workload {\color{black}demands, thus measuring or estimating the execution parameters for each of them}.
\item The introduction of a function that describes the average execution time {\color{black} of using each SEDAR strategy}.
\item A qualitative evaluation of the {\color{black} incidence of the communication pattern's influence on} temporal behavior. 
\item A discussion about {\color{black}the convenience of saving multiple checkpoints for recovery,} compared to just employing the detection mechanism. In addition, {\color{black}a brief analysis of how to determine the best moment to start protection has been conducted}.

\end{itemize}

The remainder of the paper is organized as follows: Section 2 reviews some basic concepts and related work. Section 3 describes the proposed methodology from a functional point of view, separating it in the three aforementioned strategies. Section 4 presents the evaluation of the recovery strategy, in addition to a discussion about the convenience of utilization and adaptation, considering the temporal behavior of the possible variants. Finally, in Section 5, the conclusions and future lines of work are detailed. 

\section{Background and Related Work}
\label{sec:background}

Transient faults can be classified according to their consequences on the program execution \cite{mukherjee2005soft}. 
A Latent Error (LE) is a non-harmful fault since it does not affect the final results. This kind of error alters data that are not used anymore. On the contrary, when a Detected Unrecoverable Error (DUE) occurs, a program ends suddenly; the system software can be aware of DUEs but cannot recover from them. 
In the case of a Time Out Error (TOE), the application does not finish within a stipulated time range. 
Finally, as mentioned before, when SDC takes part, the application seems to execute correctly{\color{black}, although} invalid results are produced. Particularly, in message-passing parallel programs, SDC can be sub-classified into two different types of errors according to \cite{monte2014}: (1) Transmitted Data Corruption (TDC) alters data to be sent by a process, which will propagate to others if it is not detected; (2) Final Status Corruption (FSC) has {\color{black}an} effect on non-communicated data, spreading the error in {\color{black}a} local manner and invalidating the final results.

Nowadays, there is no \textit{silver bullet} to manage frequent SDC. Some available algorithmic solutions only apply to specific kernels \cite{chen2011}{\color{black}, decreasing the cost of error detection; this kind of solution can be used in HPC environments \cite{bosilca2009algorithm}. Among them, some well-known methods like ABFT \cite{bosilca2009algorithm} can detect up to a maximum {\color{black}number} of errors in linear-algebra problems. However, each kernel requires an \textit{ad-hoc} implementation, which represents a lot of work for large HPC software{\color{black}. M}oreover, algorithm-based solutions are more intrusive as they modify the algorithms \cite{benoit2018coping}\cite{chen2011}\cite{shantharam2012fault}}. On the contrary, compiler or runtime software-based detection proposals are more general since they can be employed to any code, but at the cost of a significant increase in complexity. In addition, containment strategies seek to reduce the fault consequences, either stopping its propagation to the other nodes or to the data stored in checkpoints \cite{cappello2014}. In \cite{engel2011}, the authors increase availability and offer a trade-off between the number and the quality of components through redundancy in HPC systems. In the same way, \cite{ferreira2011} showed that replication strategy {\color{black}is} more efficient than C/R under circumstances of high error rate and {\color{black} large values of C/R} overhead.

Traditionally, SDC detection {\color{black}has been} achieved through execution replication combined with partial or total result comparison during the execution.
Software-redundancy solutions remove the need for expensive hardware performing replication at the level of threads \cite{yalcin2013}, processes \cite{shye2009} or machine status.
On the other hand, other proposals require fewer resources but at the cost of reducing {\color{black}their} accuracy. One of them is approximate replication, which implements upper and lower limits for computation results \cite{cappello2014}. 
MR-MPI \cite{engel2011} employs a transparent redundancy approach for HPC. It proposes {\color{black}a} partial process replication and can be used together with C/R in non-replicated processes \cite{ni2013}. 
rMPI \cite{ferreira2011} takes on failures by redundant execution of MPI applications. Using this protocol, the program fails only if two corresponding replicas fail{\color{black},} because each node is duplicated. The probability of simultaneous failure of a node and its replica decreases when the number of nodes increases, so redundancy scales. However, this benefit requires duplicating the number of resources and quadrupling the number of messages. 
Faults in shared-memory systems are explored in \cite{mushtaq2013}, where a scheme based on multi-threaded processes is proposed, including non-determinism management due to memory accesses.
A protocol for hybrid task-parallel MPI programs is described in \cite{martsin2015}, which carries out recovery based on uncoordinated checkpoints and message logging. Only the task that presented the error is restarted and all the MPI calls are handled inside it.
RedMPI \cite{fiala2012} is an MPI library that exploits rMPI's per-process replication to detect and correct SDC, comparing the messages sent by replicated issuers at the receiver side. 
It avoids sending all messages and comparing their entire contents through a hashing-based optimization. In addition, it does not require source code modification and guarantees determinism among replicated processes. As it offers protection even when high failure rates occur, RedMPI is a potential alternative to be used on large-scale systems.
It is shown that even a single transient error can produce deep effects on the program, causing a corruption pattern that cascades toward all other processes through MPI messages. 
In a similar way to SEDAR, RedMPI also enables replica mapping on the same physical node as the native processes, or in neighbors with lower network latency. 
Like our proposal, it monitors communications as a strategy {\color{black}which attempts to provide} the correct output.
Detection is delayed upon transmission, but, as opposed to SEDAR, validation is carried out on the receiver side. 
This produces an additional overhead, {\color{black}as well as} latency and network congestion that {\color{black}are} not present in our solution. 
Fault tolerant protocols for other parallel programming models, such as PGAS \cite{ali2011redundant} have been also explored.
The combination of checkpointing the output of tasks and replicating for application-specific detection is explored in \cite{benoit2019combining} for a linear workflow context, in the presence of both fail-stop and silent faults. {\color{black} Finally, in a recent study, the authors of \cite{benoit2019replication} explore the combination of replication with checkpointing for fail-stop errors, and compute the optimal checkpoint interval for this approach.}

\section{Description of the Methodology}

The following subsections are dedicated to the description of the basics of the different proposed options to accomplish transient fault tolerance. In addition, an evaluation of the temporal {\color{black} behavior} of implementing each specific feature is also included. In that sense, a simple model has been developed which considers the factors that have influence over the total execution time, both in the absence of faults as when a single silent error occurs during the execution. {\color{black}It is important to remark that SEDAR can {\color{black}functionally} manage multiple fault occurrence \cite{montezanti2016characterizing}{\color{black}. H}owever, the proposed {\color{black}performance model contemplates a single fault} for the sake of clarity}.

To evaluate the different strategies, a baseline is used, which consists of a manual method for ensuring reliable results. This method involves launching two simultaneous instances of the application and comparing the final results of both executions in a semi-automatic manner. In this way, the same computing resources that are consumed in our proposal are assigned to each individual instance (i.e. half of the total cores of the system){\color{black}, which is the fairest way to compare}. In the absence of faults, the final results {\color{black}will match}{\color{black}. H}owever, if a transient fault occurs, a third re-execution {\color{black}(maintaining the same mapping)} and a new comparison are required to pick the outputs of the runs that form a majority ({\color{black} using} a voting mechanism) {\color{black}as the correct ones}. 

The time elapsed by this manual method in the absence of faults (fault absence, $T_{FA}$) is given by Equation~\ref{eqn:tfa1}. {\color{black}It consists of the time spent by the two independent instances to run simultaneously ($T_{prog}$) plus the time of comparing the results of the two executions}. On the other hand, Equation~\ref{eqn:tfp1} is the time when a fault occurs (fault presence, $T_{FP}$), {\color{black}which is the time of a new re-execution and a new comparison for voting, besides a restart time ($T_{rest}$) to relaunch the third run (which takes the same time as the previous {\color{black}one because} it uses the same mapping) after the two original ones}. In Table~\ref{tab:temp-charac-params}, the parameters involved in all the equations and their meanings are summarized.

\begin{equation}
\label{eqn:tfa1}
T_{FA}=T_{prog} + T_{comp} 
\end{equation}

\begin{equation}
\label{eqn:tfp1}
T_{FP}=2(T_{prog} + T_{comp})+T_{rest}
\end{equation}

\begin{table}[ht]
\caption{Name and meaning of the parameters involved in temporal characterization of each alternative strategy} 
\centering
\begin{tabular}{| p{1.7cm}| p{7cm} |}
\hline
Parameter & \multicolumn{1}{c|}{Meaning} 
\\
\hline

$T_{prog}$     & The execution time of two instances of the original application in parallel                                                                                                   \\\hline

$T_{comp}$     & Time of semi-automatic comparison of results; may include calculating a hash \\\hline
$T_{rest}$     & Time of manually restarting the application. An automatic restart may take shorter. In simplified model, it is considered the same                              \\\hline
$f_d$        & The factor of overhead due to detection mechanism. It is application dependent and can be experimentally determined. $0<fd<$1 \\\hline                                                                    
$X$         & Instant of fault detection, expressed as a fraction of the application progress. Random $0<X<$1                                                                                             \\\hline 
$n$         & Number of checkpoints made during the whole execution, given a checkpoint interval                                                                                                           \\\hline
$t_{cs}$       & Time involved in storing a system checkpoint                                                                                                                                                                                            \\\hline
$t_i$        & Checkpoint interval.  It can be adjusted to minimize overhead                                                                                                                                                                               \\\hline 
$k$         & A number of additional checkpoints that the application needs to rollback to find a non-corrupted one. It depends on the application and the detection latency                      \\\hline 
$t_{ca}$       & The time involved in storing an application checkpoint. It should be shorter than $t_{cs}$    
\\\hline

$T_{compA}$     & The time for validating an application checkpoint; it may include calculating a hash 
\\\hline
\end{tabular}
\label{tab:temp-charac-params}
\end{table}

\subsection{Error detection with notification}

{\color{black}In order to} accomplish detection when running deterministic parallel applications, which is the first SEDAR feature, the messages between processes are validated before being sent. Thus, the error that affects any process can be isolated, preventing it from propagating to the others. 

The detection strategy consists of duplicating each application process in a thread, which requires a synchronization mechanism between both replicas. Every time {\color{black}a} communication is to be performed, the leading thread stops running and then waits for its replica to reach the same point. Once there, the detection mechanism compares the entire contents of the messages computed by both redundant threads and, if there is a coincidence, only one of them sends the message. Such a mechanism does not require additional network bandwidth. When the receiver process reaches the receive operation, it gets the message and in turn waits for its replica to synchronize. Next, it makes a copy of the received contents before resuming execution. Additionally, because a failure may have locally propagated until the end of the execution, a comparison of the final results of the application is performed, thus allowing to detect such failures. 

This detection strategy is capable of detecting failures that cause SDC {\color{black}(both TDC and FSC variants, at the cost of sacrificing latency in the latter case) and TOE}. As regards TOE, they can be detected under the assumption that the execution time of two redundant threads is similar in {\color{black}a} homogeneous, dedicated system \cite{montezanti2016characterizing}. Hence, if an appreciable delay is noticed between the two replicas, it is considered that a silent error has caused the separation of their flows. As a single time-out interval is not optimal, it should be configured taking into account the application{\color{black}'s} needs: if it is too long, the detection latency enlarges, but being too short may cause false positive detection. Anyway, a TOE is definitely detected if a process enters an infinite loop. 

\begin{figure}\centering
\subfigure[Normal operation in absence of faults]{
\includegraphics[width=0.9\columnwidth]{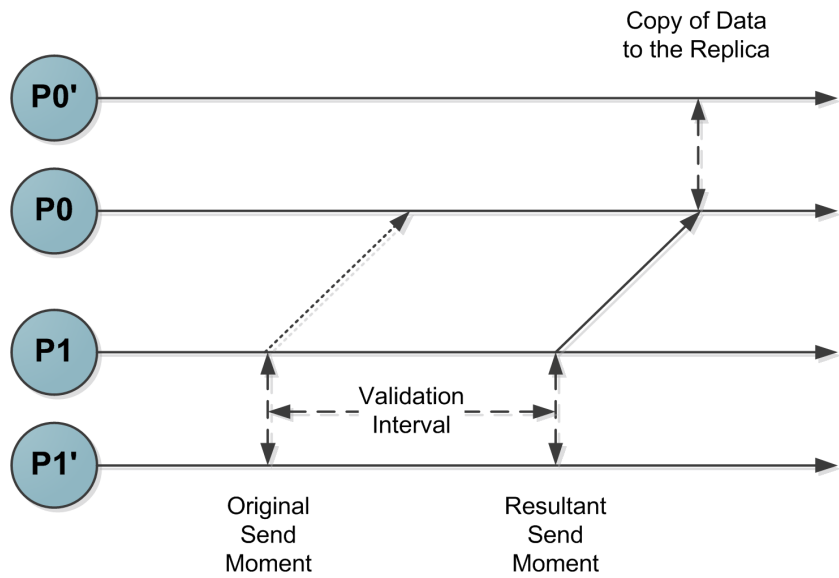}
}\\
\subfigure[Operation when detecting a fault]{
\includegraphics[width=0.9\columnwidth]{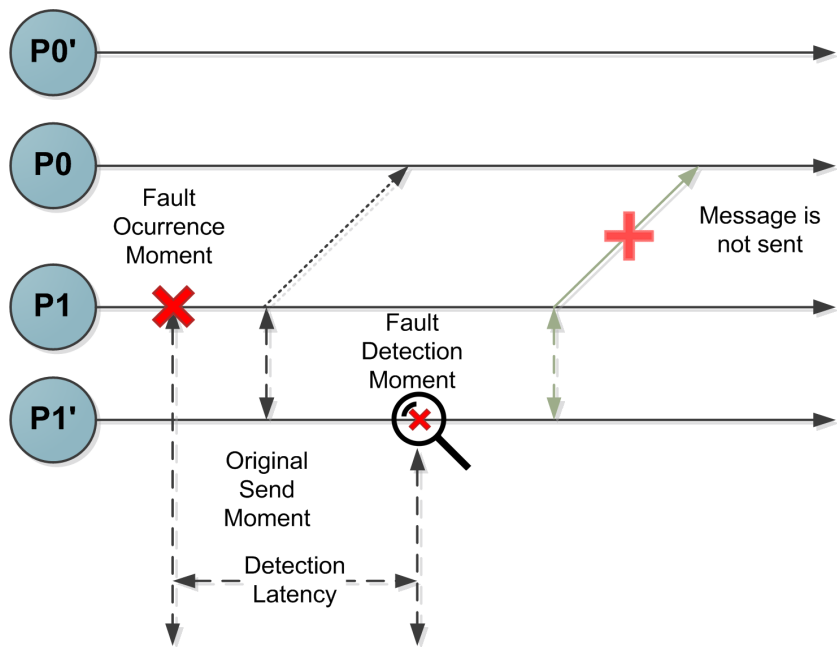}}
\caption{Outline of the operation of the detection mechanism}
\label{fig:detection}
\end{figure}

A scheme of the proposed detection strategy is shown in Figure~\ref{fig:detection}. Each replica runs in a core that shares a cache level with the core in which the original process executes{\color{black}. T}hus, the comparisons are solved with no need to access main memory, taking advantage of the memory hierarchy. It should be clear that the proposed detection mechanism is based on launching a single instance of the application, with each process internally replicated in a thread. This is different from the baseline, in which two independent instances of the application are launched in parallel{\color{black}. N}evertheless, both cases make the same use of half of the available cores from the application performance point of view.

As this methodology is based on process-replication, it is \textit{a priori} capable of performing only detection (triple redundancy needs to be used to achieve correction). The ways to accomplish recovery without the need of triplicating are described in the two next subsections. 

{\color{black} Implementing any resilience strategy involves unavoidable costs, both in execution time and in resource utilization. In particular, a {\color{black}duplication-based} mechanism aims to achieve reliability, at the cost of assigning half of the system cores to protect the executions. In this context, it is important to note that SEDAR provides fault tolerance without introducing any additional cost regarding resource utilization; it takes {\color{black}advantage} of the available cores (the intrinsic redundancy of the system), without any change or {\color{black}need} for specific hardware. As a consequence of the problems related to strong scalability, {\color{black}parallel} applications may not always make {\color{black}an} efficient use of {\color{black}all} the available cores, both in terms of time reduction and energy consumption. Therefore, {\color{black}using all the available cores may not necessarily be better (as expected) than using} half of them (some examples of this behavior can be found in~\cite{panadero2017p3s,puzyrev2015review}). {\color{black}In addition, as performance can depend on the mapping, providing resilience is a useful way to take advantage of the available resources.} 

Another remarkable aspect is that SEDAR does not modify the algorithm (as algorithm-based detectors do) and {\color{black}it} is almost transparent to the application, which makes it generally usable for message-passing applications.

The steps involved in {\color{black}the} detection, such as the replication of processes, synchronization between redundant threads, comparison before sending, {\color{black}copying} messages upon reception and final verification of the results, are the cause of the overhead introduced in the overall execution time. 

A trade-off is reached in relation to the validation interval. The overhead involved in detection is minimized if results are compared only at the end, but because the detection latency increases, a lot of computation could be useless. On the other hand, if the frequency of partial results validation is high, the introduced overhead enlarges but the fault can be quickly detected, and hence more computation is profitable. There is evidence that, depending on the computation-to-communication ratio of the particular application, as well as on the size of the workload, the overhead can differ in a significant manner \cite{monte2014}. {\color{black}As a consequence, the combination of the validation of the messages with the final comparison of results aims to detect all faults and obtain a reliable system by introducing a reasonable overhead. SEDAR's detection mechanism could also be adapted to partial replication, similar to \cite{engel2011}. However, extra work should be done to identify the application's critical parts that need to be replicated, which is a non-general procedure. Consequently, SEDAR could be disabled in the non-critical parts. Moreover, the partial replication would not result in a benefit regarding resource utilization, since the cores were previously assigned to replicas (whether they run or not). Nevertheless, it could be potentially profitable from the standpoint of energy consumption.} 

In the absence of a recovery strategy, the occurrence of an SDC or a TOE causes the {\color{black}detection-only} strategy to notify the user and lead the system to a safe stop, preventing it from delivering defective results.
As validating the messages has the effect of limiting the detection latency, the implementation of such a strategy permits {\color{black}relaunching} the execution as soon as the error is detected, thus avoiding the needless and expensive wait for the termination with corrupted results.
The execution time of the detection strategy in the absence of faults is given by Equation~\ref{eqn:tfa2}. {\color{black} The time is the same as the one of the baseline (Equation~\ref{eqn:tfa1}), but, in this case, $T_{prog}$ is negatively affected (increased) by a factor $f_d$, which represents the overhead of the detection mechanism}. If a fault occurs, the execution time is accounted in Equation~\ref{eqn:tfp2}. The first term comprises the time executed until the detection instant {\color{black}($X$) plus the whole re-execution after the stop caused by the error. Once it is detected, a restart is required, and, in the re-execution, the final comparison is needed.}\newline

\begin{equation} 
\label{eqn:tfa2}
T_{FA}=T_{prog}(1+f_d) + T_{comp}
\end{equation} 

\begin{equation}
\label{eqn:tfp2}
T_{FP}=T_{prog}(1+f_d)(X+1) +T_{rest}+ T_{comp}
\end{equation}

{\color{black}It is important to point out that the parameter $X$ represents the instant of the fault detection and not the moment of the fault occurrence (which cannot be exactly known). The value of $X$ is related to the latency of detection and depends on how the data (and then the messages) are affected by the fault, so it varies according to the communication pattern of the target application.} 

A remarkable fact is that the proposed detection strategy is equally effective when multiple errors occur {\color{black}during} the execution. The first difference, caused by an error, {\color{black}which} is observable in the contents of a message or in the final results will {\color{black}lead to the system stopping safely}. The vulnerability of this mechanism is reduced to extremely unlikely cases, which are detailed in \cite{montezanti2016characterizing,swift2005}. Consequently, despite {\color{black}the fact that} we {\color{black}have limited} the analysis of the temporal behavior to the cases of fault absence or single error occurrence, the detection strategy can handle multiple non-related errors. As regards the implementation, the required procedure for adding the detection functionality to the parallel application (that could be automated) is detailed in \cite{monte2014}. 

\subsection{Recovery based on multiple system-level checkpoints}
\label{subsec:rec-mult-syst-ckpt}
The next step in the search for transient fault tolerance consists {\color{black}of} adding a recovery mechanism. In SEDAR, it is proposed to store a chain of distributed coordinated checkpoints, built with a system-level checkpointing library.

It is not possible to ensure that any particular checkpoint holds a safe state for recovery because a silent error can spoil the internal state of one of the replicas that is {\color{black}going} to be checkpointed. Therefore, {\color{black}recovering} from the last stored checkpoint {\color{black}is not always feasible and using an} older one {\color{black}may be} required. Thus, multiple checkpoints have to be saved to guarantee recovery \cite{lu2013}. As the transient faults are fleeting, it is important to note that the restart can be attempted from the same node {\color{black}where the corruption took place}.
There are two possible cases:
\begin{enumerate}
\item \textit{The transient fault occurs and is detected inside the boundaries of a checkpoint interval}. In this situation, the last checkpoint can be used to resume the execution. As a particular case, if the detection occurs previously to the first checkpoint, the application must be relaunched from the beginning. {\color{black}This situation is outlined in Figure \ref{fig:recovery} (a).} 
\item \textit{The detection latency transposes the limits of the checkpoint interval}. This circumstance arises when the fault occurs before storing a checkpoint but the detection takes place after that. {\color{black}In this situation, the last checkpoint is} invalid, so the corresponding restart causes the same error to manifest. Consequently, the previous checkpoint must be used to attempt to {\color{black}recover}. {\color{black}Generalizing, the fault can traverse any number of checkpoints (depending on the detection latency), requiring several tries} to make rollback recovery possible. {\color{black}In turn, this situation is outlined in Figure \ref{fig:recovery} (b).} 
\end{enumerate} 

Controlled fault injection experiments are needed to verify the operation of the recovery strategy when the two aforementioned cases occur. The data corruption becomes evident as an observable difference between the memory state of the replicas{\color{black}. Hence, in order} to simulate a bit-flip in a processor register, the value of a variable is changed in only one of the replicated threads, in a single iteration of the computation. Such an injection is made from inside the code of the application. The details of the injection method are described in section~\ref{subsec:results-implem}. {\color{black} As regards the checkpoints, the best moments to take them are when the communications have just {\color{black}been} validated. This strategy reduces the windows of vulnerability \cite{swift2005} (we have previously studied this issue in~\cite{montezanti2016characterizing}), as the probability of an error corrupting the state of a checkpoint is smaller in that situation. {\color{black}However}, the mechanism {\color{black}even} works if the checkpoints are taken {\color{black}some}where, as in a more realistic scenario, with periodic checkpointing made externally to the application. However, in general, a considerable overhead would be involved if a checkpoint is taken after each communication.}

{\color{black}As {\color{black}mentioned before}, the two possible behaviors of the recovery strategy are} outlined in Figure~\ref{fig:recovery}, while Algorithm~\ref{alg:recov-system-level} describes the pseudo-code of the proposed method. For the sake of clarity, the injection mechanism is not included in the algorithm{\color{black}, nor is the recording of system checkpoints; in general, they can be taken anytime}.

\begin{figure}\centering
\subfigure[Detection latency confined within the checkpoint interval]{
\includegraphics[width=0.9\columnwidth]{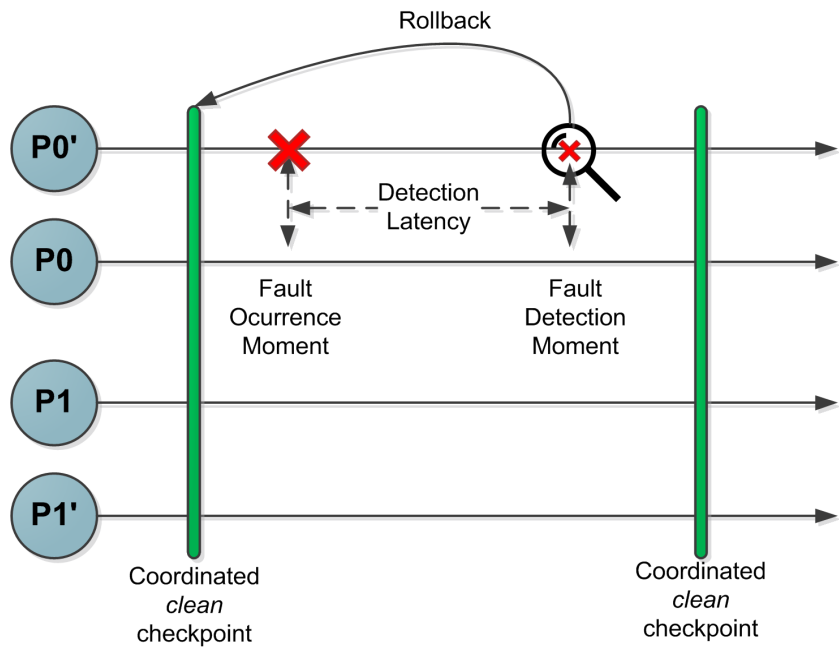}
}\\
\subfigure[Detection latency transposing the checkpoint interval]{
\includegraphics[width=0.9\columnwidth]{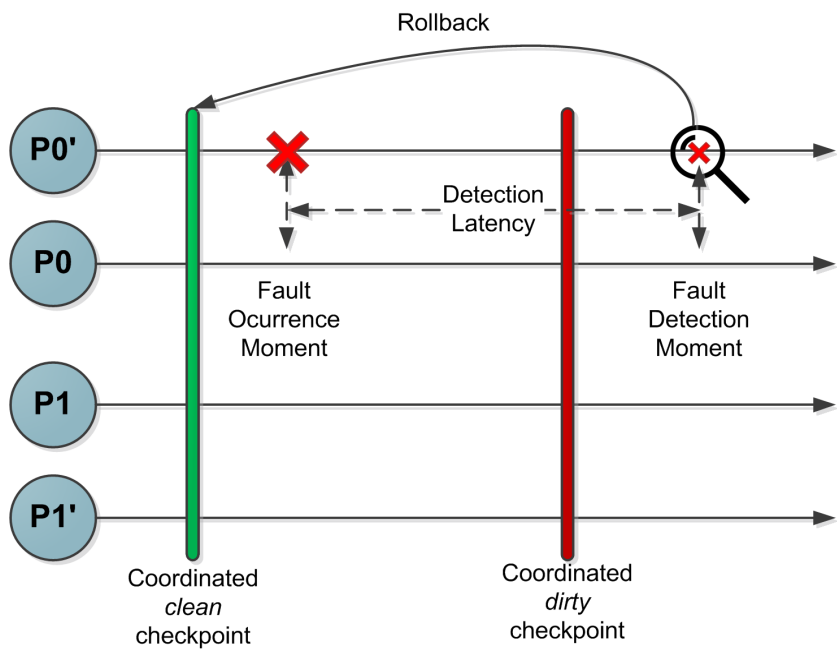}}
\caption{Possible cases of recovery using multiple system-level checkpoints, depending on the detection latency}
\label{fig:recovery}
\end{figure}

 \begin{algorithm}[h]
 {\color{black}
\caption{Recovery algorithm with multiple  system-level checkpoints}\label{alg:recov-system-level}
	\begin{algorithmic}[1]
    \CommentLine{\textit{extern\_counter} is an external counter that controls the number of rollbacks (not included in checkpoint)}	
	\State {int extern\_counter = 0;} 
    \CommentLine{\textit{fault\_detected} is a boolean variable that reports if a failure was detected in the last execution}	\State {boolean fault\_detected = FALSE;}  
    \CommentLine{run parallel application \textit{app} under SEDAR monitoring}    
    \State {SEDAR\_run(app);}  
    \CommentLine{if \textit{fault\_detected} is TRUE then a fault was detected in the last execution}    
    \While{fault\_detected == TRUE}
        \CommentLine{\textit{extern\_counter} is increased by 1}    
	    \State{extern\_counter++;} 
        \CommentLine{get the number of checkpoints done}	    
	    \State{ckpt\_count = get\_ckpt\_count();} 
        \CommentLine{calculate the number of the checkpoint for restart }	    
	    \State {ckpt\_no = ckpt\_count - extern\_counter;} 
        \CommentLine{reset detection flag for restart }	  	    
	    \State {fault\_detected = FALSE;}   
        \CommentLine{restart \textit{app} from  checkpoint \textit{ckpt\_no}}	    
	    \State {SEDAR\_restart(app, ckpt\_no);} 
	\EndWhile
	\end{algorithmic}
}
\end{algorithm}

{\color{black}Automating this method would increase its usability}. This can be accomplished if an outsider process is allowed to read the $extern\_counter$ and, based on its value, find the correct restart script to try recovery; besides that, the {\color{black} wrong-restart} checkpoint has to be erased (and stored again in re-execution). In turn, the code of fault injection is executed only once, and the target application modifies the value of $extern\_counter$ every time a fault is detected.

It is worth noting that this mechanism is also able to detect multiple faults (if they are independent of each other)~\cite{montezanti2016characterizing}. If a different fault is detected during a re-execution, the algorithm {\color{black}will} recover {\color{black}but at a sub-optimal computational cost}. In the current state of the proposal, the algorithm is {\color{black}optimized} to deal with a single error. So, when any error is detected in a re-execution, {\color{black}the algorithm assumes that {\color{black}it is the same as} was previously detected (the detection latency has exceeded} the checkpoint interval, as mentioned in Section~\ref{subsec:rec-mult-syst-ckpt}). Therefore, the detection of a different error during re-execution will generate an unnecessary rollback attempt (i.e. {\color{black}the algorithm assumes that} the last checkpoint is corrupted, although this is not the case)}. In the worst{\color{black}-case} situation, the recovery mechanism goes back to the beginning. {\color{black} While predicting the recovery time becomes difficult in the case of multiple faults, a reliable conclusion is still ensured.} The described inefficiency can be fixed by adding a more sophisticated mechanism, which is briefly described in Section~\ref{subsec:results-implem}.
 
In the absence of faults, the execution time of this mechanism is given by Equation~\ref{eqn:tfa3}. Compared with the only detection case {\color{black}(Equation~\ref{eqn:tfa2})}, the extra term accounts for the time involved in saving $n$ system-level checkpoints. When a fault occurs, the execution time is the one {\color{black}shown in} Equation~\ref{eqn:tfp3}. The parameter $k$ is the number of extra checkpoints that need to be reversed if the restart from the last one does not succeed. {\color{black}Therefore,} the third term represents the time spent in checkpointing, taking into account that various checkpoints {\color{black}($k$)} might be recorded again if {\color{black}there} {\color{black}is any} corruption that prevents successful recovery. The fourth term is an estimate of the re-execution time, considering that, on average, the fault may be detected {\color{black}midway through} the checkpoint interval. This lapse will need to be re-executed in the best case, (i.e. when recovery is possible from the last stored checkpoint ($k=0$)). If $k>0$, the same portion plus a number of checkpoint intervals (which depend on the value of $k$) will require be{\color{black}ing} re-executed. Finally, the last term represents the number of needed restarts.

\begin{equation}
\label{eqn:tfa3}
T_{FA}=T_{prog}(1+f_d) + T_{comp} + nt_{cs}
\end{equation}

\begin{equation}
\label{eqn:tfp3}
\begin{split}
T_{FP}=T_{prog}(1+f_d) + T_{comp} + (n+k)t_{cs} +\\+ (\sum_{m=0}^{k}(k-m+1/2))t_i+(k+1)T_{rest}
\end{split}
\end{equation}

When system-level checkpoints are the only available option, this strategy becomes appropriate, despite having two significant limitations. The first one refers to the amount of required storage. None of the checkpoints can be erased, given the uncertainty about the validity of the data recorded in them: if a consistent checkpoint cannot be found, a significant part of the application will need to be re-executed; in an extreme case, the whole execution will have to be relaunched from the beginning \cite{lu2013}. In any case, the negative impact of multiple checkpoints over the storage can be reduced by solutions based on multi-level checkpointing \cite{cappello2014}. The second important drawback is related to scalability: in upcoming exascale systems, and despite some considerable efforts \cite{cao2016system}, coordinated-system-level checkpoints would not be the most suitable solution, because they keep a large amount of information related to the system. In an unrefined version, our method is an expensive approach, because it needs to keep an undetermined number of active checkpoints and may require several restart attempts. Instead, user-level checkpoints are becoming more usual, especially due to their lower costs and portability {\color{black}options}~\cite{martsin2015}.

\subsection{Recovery based on a single safe application-level checkpoint}
This third alternative included in SEDAR is designed to overcome the limitations caused by the utilization of system-level checkpoints. In this context, and despite requiring detailed knowledge of the application{\color{black}'s} internal organization (computing and communication), user-level checkpoints are a more appropriate option, given the fact that they only save the application-related information \cite{benoit2018coping}. Besides that, they are smaller, more portable and scale better than the system-level versions. As a consequence, the utilization of a single user-level checkpoint for recovery is proposed, in conjunction with a strategy to ensure the validity of the last recorded checkpoint. Therefore, the prior checkpoints can be removed, thus decreasing storage usage and reducing the relaunching latency.

The proposed solution {\color{black}is based on recording} per-thread user-level checkpoints, taking advantage of the synchronization mechanism between replicas. Such checkpoints {\color{black}just save} the set of variables that are significant to the application at that specific moment. As both thread checkpoints are stored, a hash on each one is calculated. {\color{black}To collate the two hashes, the mechanism used for validating message contents in the detection phase is employed again. Hence,} the checkpoint is considered \textit{valid} only if the comparison proves {\color{black}to be} true. In this situation, the previous checkpoint can be safely discarded to save storage, as the current one constitutes a consistent state for recovery. 
On the other hand, if a difference is detected on the verification phase, it is necessarily due to a fault {\color{black}that has} occurred within the last checkpoint interval. As this checkpoint is considered \textit{corrupted}, it is not possible to use it for recovery and {\color{black}it} should be erased. Then, the execution has to be resumed from the previous checkpoint. As a consequence, there is a single valid checkpoint at any given time (except for the validation interval), {\color{black}which is independent from} the comparison result. Algorithm~\ref{alg:recov-user-level} describes the pseudo-code of the proposed mechanism.

\begin{algorithm}[h]
\caption{Recovery algorithm with application-level checkpoints}\label{alg:recov-user-level}
	\begin{algorithmic}[1]
    \Function {usr\_ckpt}{$n$}  \CommentSingleLine {usr\_ckpt function definition}     
	    \For {(tid=0; tid $<$ 2; tid++)} \CommentSingleLine{for both replicas}
	        \CommentLine {record its custom checkpoint}
		    \State {store\_all\_significant\_variables(tid);} 
		    \State {hash\_array[tid]=compute\_hash(tid);}
	    \EndFor
	    \State {synch\_threads();} \CommentSingleLine {wait for each other}
	    \CommentLine {only one of the replicas compares hashes}
	    \If {tid==0} 
		    \If {hash\_array[0]==hash\_array[1]} \CommentSingleLine { they match }
		    \CommentLine {delete own checkpoint}
			    \State{remove\_all\_significant\_variables(tid);} 
                \CommentLine{this is a valid checkpoint so the previous can be discarded}			    
			    \State \Return TRUE;	
			\Else \State{}
			    \Return {FALSE} \CommentSingleLine{ this is a corrupted checkpoint}
			   \EndIf
	    \EndIf
	\EndFunction
    \State{}
    \State {(...)} \CommentSingleLine{ application code }
    \CommentLine {\textit{n} represents the current checkpoint}
    \If {usr\_ckpt(n)== TRUE}	
    \CommentLine{delete previous checkpoint since the current is valid}
	    \State {remove\_usr\_ckpt(n-1);}  
    \Else    
    \CommentLine{ remove current corrupted checkpoint}
	    \State{remove\_usr\_ckpt(n);} 
	    \CommentLine{ restart from previous checkpoint}
	    \State{restart\_from\_usr\_checkpoint(n-1);} 
	\EndIf
	\end{algorithmic}
\end{algorithm}

In the absence of faults, the execution time of this mechanism is given by Equation~\ref{eqn:tfa4}. The time is equal to the {\color{black}detection-only} strategy, but incorporates an additional {\color{black}last} term which represents the time employed for $n$ user-level checkpoints to be recorded {\color{black}($t_{ca}$) after being validated ($t_{compA}$)}. Instead, Equation~\ref{eqn:tfp4} accounts the time when a fault occurs. The fourth {\color{black}added} term shows that, on average, just half of the checkpoint interval has to be re-executed, as barely a single rollback is required. {\color{black}For the sake of clarity: as each checkpoint is validated, the latency of detection is confined within the checkpoint interval. In the worst{\color{black}-case scenario}, the re-execution time will be $t_{i}$, if the error is detected just before taking a new checkpoint; while, in the best{\color{black}-case scenario}, in which the error is detected as soon as a checkpoint has been taken, it will be near to 0. As the probability of an error is equally distributed along with the checkpoint interval, we state that, {\color{black}on} average, the re-execution time is (1/2) $t_{i}$}. {\color{black}Ultimately, the last term represents the only restart time that is required, as the algorithm performs a single rollback at most}. It is important to notice that $T_{comp}$ represents the time required for the validation of application results, while $T_{compA}$ comprises the time required to validate an application-level checkpoint.

\begin{equation}
\label{eqn:tfa4}
T_{FA}=T_{prog}(1+f_d) + T_{comp} + n(t_{ca} +T_{compA})
\end{equation}
\begin{equation}
\label{eqn:tfp4}
\begin{split}
T_{FP}=T_{prog}(1+f_d) + T_{comp} + n(t_{ca} +T_{compA}) +\\+ (1/2)t_i+T_{rest}
\end{split}
\end{equation}

{\color{black}\subsection{Average Execution Time}

As {\color{black}previously mentioned}, Equations~\ref{eqn:tfa2} to~\ref{eqn:tfp4} describe the time {\color{black}required} by each strategy {\color{black}in two cases: (1) in the absence of faults ($T_{FA}$); and (2) in the presence of a single silent fault ($T_{FP}$)}. As {\color{black}a fault has an associated occurrence probability}, we introduce a general formulation that {\color{black}predicts} the Average Execution Time considering it and, as a consequence, the {\color{black}\textit{Mean Time Between Errors} (MTBE)} parameter. This function allows us {\color{black}to estimate} the average overhead introduced by each SEDAR strategy.
Let {$\alpha$} be the probability of a fault occurrence. Then, the Average Execution Time function is given by:

\begin{equation}
\label{eqn:aet1}
AET=T_{FP}(\alpha) + T_{FA}(1-\alpha)
\end{equation}

The $MTBE$ of a system with $N$ processors decreases linearly with $N$, that is {$MTBE$ = $MTBE_{ind}$/$N$}, where $MTBE_{ind}$ is the $MTBE$ of an individual processor \cite{benoit2018coping}. If $\lambda$ = 1/$MTBE_{ind}$ is the silent error rate of an individual processor, {\color{black}then} the silent error rate of the whole system is $\lambda N$. {\color{black}Assuming that errors occur according to an exponential distribution}, the probability of a silent error affecting a computation that lasts $T_{prog}$ and executes on a system with $N$ processors is: 

\begin{equation}
P(N,T_{prog}) = 1 - e^{-\lambda NT_{prog}} = 1 - e^{-NT_{prog}/MTBE_{ind}} = 1 - e^{-T_{prog}/MTBE}
\end{equation}

The latter expression is the probability of a silent fault occurrence, that is, $\alpha$. {\color{black}By} considering this in Equation~\ref{eqn:aet1}, we can obtain the Average Execution Time as a function of the $MTBE$ of the system and the baseline program execution time $T_{prog}$.

\begin{equation}
\label{eqn:aet2}
AET=T_{FP}(1 - e^{-T_{prog}/MTBE}) + T_{FA}(e^{-T_{prog}/MTBE})
\end{equation}

Equation~\ref{eqn:aet2} is useful {\color{black}for estimating} the average overhead of using each SEDAR strategy, considering both the cases of fault absence and fault presence. If Error Detection with Notification strategy is used, $T_{FA}$ and $T_{FP}$ of Equation~\ref{eqn:aet2} are obtained from Equations~\ref{eqn:tfa2} and~\ref{eqn:tfp2}; when using Recovery based on Multiple System-Level Checkpoints, they are obtained from Equations~\ref{eqn:tfa3} and~\ref{eqn:tfp3}; and when Recovery based on a Single Safe Application-Level Checkpointing is the chosen {\color{black}strategy}, the times are obtained from Equations~\ref{eqn:tfa4} and~\ref{eqn:tfp4}.
}

\section{An Evaluation of the Recovery Method}

\subsection{Analytical Model}

{\color{black}To describe and validate the functional behavior of the detection and automatic recovery strategies in the presence of faults, we have built a model based on combining a well-known test application with a complete, controlled workfault. The analytical model contemplates all the possible faults that can occur}, based on the deep knowledge of the behavior of the application. Each fault has a predictable effect, a moment in which it is certainly detected and a determinable point for recovery. Obviously, there are infinite physical possibilities of fault occurrences, but all of them are represented in the listed {\color{black}scenarios; a scenario represents a class of errors, so it 
includes a set of cases that behave in the same way}. For each experiment, a single fault is injected.

The test application is synthetic, built-up over an MPI Master/Worker matrix multiplication ($C=A \times B$). The modifications consist of replicating processes in threads for detection, in addition to final validation of the resulting matrix. Every time the application performs a communication, a system-level checkpoint is {\color{black}carried out as} messages are sent only if the involved data are safe. The matrix-multiplication has been selected because it is a regular, computationally-intensive, representative parallel application, with a well-known communication pattern. The deep knowledge about its behavior allows the clear identification of the moments of communication between processes and of the data involved in each communication. As a consequence, the precise effect of each injected fault can be predicted, as well as the state of each recorded checkpoint (\textit{clean} or \textit{dirty}), and, therefore, which checkpoint makes the recovery possible.

The pseudo-code for the test application is shown {\color{black}in} Algorithm~\ref{alg:test-app}. 

\begin{algorithm}[h]
\caption{Pseudo-code for the test application}\label{alg:test-app}
	\begin{algorithmic}[1]
		\State{ SEDAR\_Init()} 
		\State{ SEDAR\_Ckpt()} \CommentSingleLine{Checkpoint \#0 (CK0)}
		\State{ SEDAR\_Scatter(A)} \CommentSingleLine{Master scatters matrix A (SCATTER)}
		\State{ SEDAR\_Ckpt()} \CommentSingleLine{Checkpoint \#1 (CK1)}
		\State{ SEDAR\_Bcast(B)} \CommentSingleLine{Master broadcasts matrix B (BCAST)}
		\State{ SEDAR\_Ckpt()} \CommentSingleLine{Checkpoint \#2 (CK2)}
		\State{ matmul(A,B,C)} \CommentSingleLine{Each process computes its block (MATMUL)}
		\State{ SEDAR\_Gather(C)} \CommentSingleLine{Master gathers matrix C (GATHER)}
		\State{ SEDAR\_Ckpt()} \CommentSingleLine{Checkpoint \#3 (CK3)}
		\If{ rank == MASTER}
		\CommentLine{Master validates final result (VALIDATE)}
		    \State{ SEDAR\_Validate(C)}
		\EndIf
	\end{algorithmic}
\end{algorithm}

The possible scenarios for injection experiments are organized according to the following criteria:

\begin{itemize}
\item $P_{inj}$: the execution instant where the injection is carried out, taking as reference the structure of the application (e.g. between the SCATTER and CK1).
\item $Process$: if the injection is made in the code executed by the Master or any Worker.
\item $Data$: if the injection is made either on an element of the matrix $A$, $B$ or $C$, or on an index variable. Also, if the injected value is used by the Master or a Worker for its computation (e.g. A(M), C(W), i(M), etc).
\item $Effect$: TDC, FSC, LE or TOE.
\item $P_{det}$: the execution moment where the fault is detected. It could be at communication or at the final validation.
\item $P_{rec}$: the nearest checkpoint from which it is possible to recover.
\item $N_{roll}$: the number of attempts required for correct recovery. The possible values are: 0, if the injected fault causes a LE; 1, if recovery is possible from the last recorded checkpoint; 2, if it is necessary to rewind to the last but one checkpoint; and so on. 
\end{itemize}

Based on the combinations of these factors, we have designed a set of 64 injection experiments {\color{black}that} cover all the situations that can occur in the target test application. The 64 scenarios have been designed according to the following criteria: the faults are injected both in the code executed by the Master and by the Workers, in elements of each of the three matrices. The injections in the Master code are made both in data that it transmits and in other{\color{black}s} that are kept for local use. The {\color{black}injections} in the Workers' code are made in data that will be transmitted, as the Workers do not locally retain results. In both the Master and the Workers, there are faults injected after each checkpoint, i.e. the checkpoint is \textit{clean}, so recovery is possible{\color{black}. Nevertheless}, other faults are injected after a communication operation but before the subsequent checkpoint (i.e. between them, into already validated data), making that checkpoint \textit{dirty} and forcing more than one rollback to recover. On the other hand, injections in index variables are made, both in the Master and the Workers codes, during the actual matrix-multiplication operation, in order to make the processing time of both redundant threads asymmetric. As {\color{black}previously mentioned}, every fault affecting a certain subset of data, and occurring (at any moment) during the lapse of the execution comprised in a particular scenario, is detected at the same time, so it can be recovered from the same checkpoint. In other words, any case of error has a similar effect to one of the 64 provided scenarios; these 64 scenarios are derived from the study of the application behavior.

To illustrate the method followed, only a few representative scenarios 
are detailed in Table~\ref{tab:inject-cases}. 
These scenarios were selected mainly to make evident the four possible effects of a fault (TDC, FSC, LE or TOE), but also to display injections made in the codes of both the Master and the Workers, showing different moments of detection and reflecting various possible situations of recovery (i.e. no need to rollback, rollback to the last checkpoint or multiple rollback attempts).

\begin{table}[ht]
\caption{Selected representative injection scenarios: effects and predicted points of detection and recovery}
  \begin{adjustbox}{width=\textwidth}
\begin{tabular}{|c|c|c|c|c|c|c|c|}
\hline
\textbf{Scenario} & \textbf{$P_{inj}$} & \textbf{Process} & \textbf{Data} & \textbf{Effect} & \textbf{$P_{det}$} & \textbf{$P_{rec}$} & \textbf{$N_{roll}$} \\ \hline
2             & CK0 - SCATTER & Master           & A(W)          & TDC             & SCATTER       & CK0           & 1              \\ \hline
29             & BCAST - CK2   & Worker           & C(W)          & LE              & -             & -             & 0              \\ \hline
50             & GATHER - CK3  & Master           & C(M)          & FSC             & VALIDATE      & CK2           & 2              \\ \hline
59             & MATMUL        & Worker           & i(W)          & TOE             & GATHER        & CK2           & 1              \\ \hline
\end{tabular}

  \end{adjustbox}
\label{tab:inject-cases}
\end{table}

From Table~\ref{tab:inject-cases} we can observe that:

\begin{itemize}
\item In Scenario 2, the injection is carried out modifying matrix $A$ from the Master process between CK0 and SCATTER stages. The injected element of $A$ is going to be transmitted to a Worker, so this injection produces a TDC error, which will be detected at SCATTER moment. The recovery is possible from the last checkpoint {\color{black}carried out} (CK0, a \textit{clean} checkpoint).
\item An example of LE error is described in Scenario 29. The injection happens between BCAST and CK2 affecting matrix $C$ from a Worker. As this matrix has not been computed yet and will be overwritten afterward{\color{black}s}, the error does not modify the final result. 
\item Scenario 50 shows the details for an injection experiment that causes a FSC error. The injection is made in an element of matrix $C$ that has {\color{black}already} been calculated and received by the Master (GATHER), but before making the checkpoint CK3. The error will be detected at the VALIDATE stage, and, because CK3 is \textit{dirty} (the fault occurred before recording it), an additional rollback is required to recover.
\item A possible TOE error is detailed in Scenario 59. This injection takes place at MATMUL stage and affects an index variable used by a Worker, causing one of the replicas to restart its computation after it has already done part of its task. This causes a delay in the affected thread, which is detected as a TOE; only the other replica reaches the GATHER operation within a configurable lapse. The recovery is built from the CK2 point (a single rollback is enough). 
\end{itemize}

A conclusion of our analysis is that any random fault that can occur along the execution resembles some of the modeled scenarios. A method of analysis like the one described, based on the knowledge of the target application and the moments of checkpointing, in combination with controlled and systematic fault injection, allows us to predict the behavior of both the detection and the automatic recovery mechanisms. Thus, the efficiency of the strategy is shown when running in a real environment with random faults.

It is to be noted that {\color{black}the aim of the performed functional analysis is the} evaluation of the efficacy of SEDAR's detection and recovery mechanisms. As the correct operation of both mechanisms is {\color{black}checked} for all the errors that can occur in the particular selected test application, and other types of errors {\color{black}do not exist}, the carried-out validation verifies the functional suitability of SEDAR.

\subsection{Results over a Real Implementation}
\label{subsec:results-implem}
{\color{black}All the experiments that are described in this section have been carried out utilizing standard tools}. 
The implementation of the fault tolerance strategy consists of a library of modified MPI functions and data types with extended functionality for fault detection. This includes {\color{black}a} buffers comparison before sending, message copies upon reception, and synchronization between replicated threads. 

The coordinated system-level checkpoints are built with {\color{black}the} DMTCP library~\cite{ansel2009dmtcp}, which generates distributed-per-process checkpoint files, and a single restart script for each checkpoint. All processes of the application call SEDAR\_Ckpt(), but a single process (e.g. the Master) is in charge of checkpointing the whole application. Both redundant threads of {\color{black}this} process synchronize with each other (in the same way that they do when a message is to be sent), and only one of them calls DMTCP\_Ckpt() from inside of SEDAR.

The fault injection is made from inside the test application. An \textit{ad-hoc} function is included in the library, {\color{black}which} contains the 64 scenarios, and {\color{black}a} conditional compilation is used to make a single injection in each experiment. Depending on the number of the particular injection scenario, the function is invoked in a different place during the execution.
This function works in conjunction with a file that is used to control if an injection has been already made (named \textit{injected.txt}). The content of this file is evaluated in each function call. {\color{black}In the first instance}, the file contains the value \textit{0} and when the injection is made, its content is \textit{incremented} (it is changed to \textit{1}). In the recovery process, the code is re-executed calling the injection routine again. As the file content is re-evaluated, the function returns without making a new injection since it has changed. This flag needs to be external to the application so that its content is independent of the checkpointing storage (i.e. when a checkpoint is made, the value in the file is {\color{black}not} affected). 

The recovery mechanism, based on keeping a chain of checkpoints, uses another file to control how many times the same fault has been detected (named \textit{failures.txt}). This file is initialized with \textit{0} and its content is incremented each time that a fault is detected. {\color{black}Assuming} that a single error occurs during execution, {\color{black}then} the file contains \textit{1} for the first detection, and a rollback to last restart script is tried. If an error is detected during re-execution, it is assumed that it is the same error as the previous one {\color{black}and} that the checkpoint is \textit{dirty} due to the detection latency. In this case, the file's content becomes \textit{2} and a rollback to the prior checkpoint is {\color{black}attempted}. Therefore, the content of the file is used to choose the number of the restart script that has to be executed. {\color{black}Once a}gain, the file needs to be external to the application, so that its content is independent of the checkpoint storage.

{\color{black} To recover from multiple different faults, the mechanism should be slightly modified to achieve a better performance}. Some additional data, related to the current fault, {\color{black} might also need to be} stored in the file, in order to {\color{black}be able to distinguish between a repetition of the previous fault and a new fault}. In {\color{black}the latter case, the file size would not increase and the mechanism would behave as in the "first" detection scenario}. 

The 64 injection experiments{\color{black}, each one being representative of a particular scenario,} were performed over the test application, using {\color{black}two nodes of a Blade cluster with eight nodes}. Each node contains two quad-core Intel Xeon e5405 2.0GHz processors {\color{black}(6MB L2 cache, shared between pairs of cores)}, 10GB RAM memory (shared between both processors) and 250GB local disk storage. {\color{black}The operating system is GNU/Linux Debian 6.0.7 (64 bits, kernel version 2.6.32), the message-passing library is MPICH (version 3.3.1), and the checkpointing library is DMTCP (version 2.4.4)}.

\begin{figure}[ht]
    \centering
    \includegraphics[width=1\textwidth]{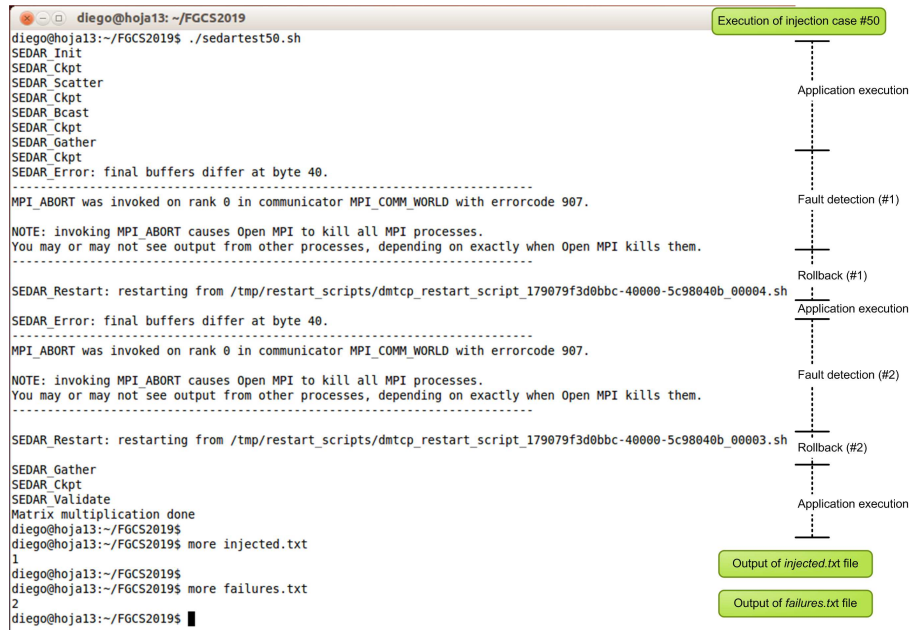} 
    \caption{\label{fig:inject-output} Output of the execution of the test application when the fault of Scenario 50 is injected}
\end{figure}

Figure~\ref{fig:inject-output} shows the output file of one of the injection experiments, namely Injection Scenario 50, in order to demonstrate the followed methodology and the obtained behavior.  

{\color{black}In order to provide more} detail, {\color{black}in this stage of {\color{black}SEDAR's} functional validation,} our implementation of fault-tolerant MPI functions is based on point-to-point communications, which make{\color{black}s it} more general and allow{\color{black}s} us to show up the FSC scenarios, at the expense of sacrificing performance. However, we {\color{black}have} also developed versions of {\color{black}optimized} collective communications, which are {\color{black} used in the stage of the temporal behavior {\color{black}evaluation}}. It is important to note that in collective communications, the sender process also participates in {\color{black}these}, either sending or receiving. If our test application just uses collective operations, the corrupted data gets transmitted and hence it is validated. In this way, only TDC scenarios remain and FSC scenarios should not be {\color{black} present} any longer. It is clear that this idea could be extended to many other applications.

\subsection{Evaluation of the Temporal {\color{black}Behavior}}
\label{sec:eval-temp-costs}
To show how our {\color{black} descriptive} model can be used to evaluate the temporal behavior of each alternative, a {\color{black}set of simple examples is presented, which includes real measured values for the parameters in Table~\ref{tab:temp-charac-params}, taken from carried-out tests}. 

{\color{black}To enrich the analysis}, we have used three parallel benchmarks for the testing stage: matrix multiplication; Jacobi's method for Laplace's equation \cite{jacobiAndrews}; and DNA sequence alignment with Smith-Waterman algorithm \cite{rucci2011pdpta}. These applications have been selected because they are well-known, computationally intensive, and representative of scientific computing. In addition, they allow {\color{black}us} to study the effects of having different communication patterns: Master-Worker, Single-Program-Multiple-Data (SPMD) and Pipeline, respectively \cite{monte2014}. {\color{black}The comparison between execution times of the three target applications, between raw MPI versions and our MPI-based implementation of SEDAR, has the aim of measuring the execution parameters with each application and pointing out the differences {\color{black}according} to the distinct communication patterns. The SEDAR implementation} includes mechanisms for duplicating processes, synchronizing replicas, comparing and copying the messages' contents, and validating the final results. Each experiment has been repeated five times, and average values have been taken.

 The parameter $f_d$, the detection overhead, was obtained by comparing the execution time of the SEDAR detection mechanism, in {\color{black}the} absence of faults, with the time of executing the manual detection strategy (baseline), for each target application. Explicitly:

\begin{equation}
\label{eqn:fd}
f_d = \frac{T_{SEDAR\_det\_FA} - (T_{prog}+T_{comp})}{T_{prog}+T_{comp}}
\end{equation}\newline

where ${T_{SEDAR\_det\_FA}}$ is the time of Equation~\ref{eqn:tfa2} and ($T_{prog}+T_{comp}$) is the time of Equation~\ref{eqn:tfa1}.

As regards the parameter $T_{comp}$, it has been measured for a manually-launched program that compares the contents of two binary files. It is similar to the one {\color{black}required} by the function SEDAR\_Validate().
{\color{black}An average value of the parameter $t_{cs}$ was obtained by measuring with tools provided by DMTCP library. As regards $T_{rest}$, it has been indirectly measured (by fault injection experiments which demand recovery), but the obtained values were consistent to the assumption that the time needed to perform a checkpoint can be considered equal to the time needed to load a checkpoint from storage} \cite{fialho2011missing}. 


{\color{black} To achieve long-lasting executions for the three applications (around 10 hours for the baseline case), we have adjusted the execution parameters. The matrix product has been repeated 100 times using N = 8192. For the Jacobi algorithm, the size of the workload was N = 8192 and the number of iterations I = 300.000. Finally, for DNA sequence alignment, the length of the sequences was set to N = 2^$22$ = 4.194.304. In this context, the parameter $t_i$ has been fixed to 1 hour for all the experiments. Regarding $n$, {\color{black}this} is the number of checkpoints that have been recorded with such an interval value{\color{black}, and} it is obtained by dividing the time of the only detection strategy (Equation~\ref{eqn:tfa2}) by the checkpoint interval $t_i$. 
It is worth noting that (instead of being arbitrarily assigned) the checkpoint interval can be determined, for example, with Daly's formula \cite{daly2006higher}, which takes into account the $MTBE$. The checkpoint interval is intended to be a trade-off between maintaining a low overhead, due to checkpointing, and a reasonable time of rework if a fault occurs.

The parameter $X$, which depends on the detection latency, has been manually assigned for three cases.} On its behalf, the parameter $t_{ca}$ was estimated assuming that {\color{black} an application-level checkpoint is more lightweight for storing} than its system-level counterpart. {\color{black} Last, $T_{compA}$ (the validation time of an application-level checkpoint) has been estimated to be equal to the time of validating the results (i.e. parameter $T_{comp}$), as a way to simplify the model}.

Table~\ref{tab:temp-ev-values} contains the list of all values used for the temporal evaluation, obtained as described, whereas Table~\ref{tab:temp-ev-results} shows the resulting times in each case.

{\color{black}It may be recalled that in the manual strategy (execution of two simultaneous MPI independent instances), a total of eight MPI processes have been launched, with a maximum of four processes mapped in each node, which means that just four cores have been used in that node. The same mapping was assigned in SEDAR implementation, but in this case, all the cores of each node have been used, as the redundant threads run on free cores.}

An analysis of the data contained in Table~\ref{tab:temp-ev-values} reveals some remarkable facts. In all the cases, the detection overhead $f_d$ is very low{\color{black}. H}owever, the biggest value is for Jacobi's method, which is the application with the most frequent communications. This is the expected behavior since $f_d$ is tightly associated with messages. On the other hand, the matrix product, which is computation bounded, presents negligible detection overhead. 

On the other hand, $t_{cs}$ is directly related with the size of the workload $W$ for each application, and therefore with the amount of memory spent, as can be seen in Table~\ref{tab:temp-ev-values}. The matrix product is the most memory-consuming application. The Master process handles the three entire matrices, whereas each Worker handles the entire B matrix, plus its corresponding chunks of A and C. As regards Jacobi's method, one of the processes handles the two entire matrices, whereas the other processes handle their corresponding chunks of them. Finally, in Smith-Waterman algorithm, all the processes require local buffers; one of them handles two entire sequences, whereas the others handle one entire sequence plus their corresponding chunk of the other. It may be recalled that for the three benchmarks, all the processes are replicated.
 
%

{\color{black}Finally}, $T_{comp}$ is associated with the size of the results that have to be validated. In the matrix product, the entire matrix C is validated for each instance, so the obtained value is significant compared to the other applications. On the other hand, in the DNA sequence alignment, only the similarity score has to be validated, which involves negligible time. As a middle ground, in Jacobi's method, a single matrix needs validation.


\begin{table} [h]
\caption{Values of the parameters utilized in the temporal evaluation of each alternative strategy}
\centering
{\color{black}

\begin{tabular}{|c|c|c|c|}  
\hline
Parameter & {MATMUL} & {JACOBI} & {SW} 
\\
\hline

$T_{prog}$ [hs]     
& \begin{tabular}[c]{@{}l@{}}10.21\end{tabular} 
& \begin{tabular}[c]{@{}l@{}}8.92\end{tabular} 
& \begin{tabular}[c]{@{}l@{}}11.15\end{tabular}                \\\hline
$T_{comp}$ [s]     
& \begin{tabular}[c]{@{}l@{}}42 \end{tabular}
& \begin{tabular}[c]{@{}l@{}}1\end{tabular}
& \begin{tabular}[c]{@{}l@{}}$<$1\end{tabular}
\\\hline
$f_d$ [$\%$]       
& \begin{tabular}[c]{@{}l@{}}$<$0.01\end{tabular}                   & \begin{tabular}[c]{@{}l@{}}0.6\end{tabular}     
& \begin{tabular}[c]{@{}l@{}}0.05\end{tabular}     
\\\hline

$X_1$; $X_2$; $X_3$ [\%] 
& \multicolumn{3}{c|}{30; 50; 80}
\\\hline 
 
$t_i$ [hs] &
 \multicolumn{3}{c|}{1}
\\\hline
$n$
& \begin{tabular}[c]{@{}l@{}}10\end{tabular}   
& \begin{tabular}[c]{@{}l@{}}8\end{tabular}   
& \begin{tabular}[c]{@{}l@{}}11\end{tabular}   
\\\hline 

$W$ [MB]
& \begin{tabular}[c]{@{}l@{}}6016\end{tabular}   
& \begin{tabular}[c]{@{}l@{}}1920\end{tabular}   
& \begin{tabular}[c]{@{}l@{}}152\end{tabular}   
\\\hline 
$t_{cs}$ [s] 
& \begin{tabular}[c]{@{}l@{}}14.10\end{tabular}   
& \begin{tabular}[c]{@{}l@{}}9.62\end{tabular}   
& \begin{tabular}[c]{@{}l@{}}2.55\end{tabular}   
\\\hline
$T_{rest}$ [s]  
& \begin{tabular}[c]{@{}l@{}}14.10\end{tabular}   
& \begin{tabular}[c]{@{}l@{}}9.62\end{tabular}   
& \begin{tabular}[c]{@{}l@{}}2.55\end{tabular}   
\\\hline
$t_{ca}$ [s] & 10.58 & 9.11 & 1.92 
\\\hline
$t_{compA}$ [s] & 42 & 1 & $<$1                 
\\\hline 
\end{tabular}
}

\label{tab:temp-ev-values}
\end{table}

\begin{table} [h]
\caption{{\color{black}Execution times [hs]} of all SEDAR alternative strategies, both in the absence and in the presence of faults, compared with the baseline}
\centering
{\color{black}
  \begin{adjustbox}{width=\textwidth}
\begin{tabular}{|c|c|c|c|c|}  
\hline
$\#$ & Situation & MATMUL & JACOBI & SW
\\\hline\hline 

1  & Baseline, without fault (Eq. 1)      
& \begin{tabular}[c]{@{}l@{}}10.22\end{tabular}                   
& \begin{tabular}[c]{@{}l@{}}8.92\end{tabular}                      & \begin{tabular}[c]{@{}l@{}}11.15\end{tabular}                      \\\hline
2 & Baseline, with fault (Eq. 2)     
& \begin{tabular}[c]{@{}l@{}}20.45\end{tabular} 
& \begin{tabular}[c]{@{}l@{}}17.85\end{tabular} 
& \begin{tabular}[c]{@{}l@{}}22.35\end{tabular} 
\\\hline\hline 
3 & Only detection, without fault (Eq. 3)     
& \begin{tabular}[c]{@{}l@{}}10.23\end{tabular}                  
& \begin{tabular}[c]{@{}l@{}}8.97\end{tabular}                  
& \begin{tabular}[c]{@{}l@{}}11.16\end{tabular}                  \\\hline
4 & Only detection, with fault (Eq. 4, X = 30\%)        
& \begin{tabular}[c]{@{}l@{}}13.29\end{tabular}
& \begin{tabular}[c]{@{}l@{}}11.67\end{tabular}                 
& \begin{tabular}[c]{@{}l@{}}14.50\end{tabular}                 \\\hline
5 & Only detection, with fault (Eq. 4, X = 50\%)        
& \begin{tabular}[c]{@{}l@{}}15.33\end{tabular}
& \begin{tabular}[c]{@{}l@{}}13.46\end{tabular}                 
& \begin{tabular}[c]{@{}l@{}}16.73\end{tabular}                 \\\hline
6 & Only detection, with fault (Eq. 4, X = 80\%)        
& \begin{tabular}[c]{@{}l@{}}18.39\end{tabular}
& \begin{tabular}[c]{@{}l@{}}16.16\end{tabular}                 
& \begin{tabular}[c]{@{}l@{}}20.08\end{tabular}                 \\\hline \hline 
7 & Multiple checkpoints, without fault (Eq. 5)
& \begin{tabular}[c]{@{}l@{}}10.26\end{tabular}                    
& \begin{tabular}[c]{@{}l@{}}9.00\end{tabular}                    
& \begin{tabular}[c]{@{}l@{}}11.17\end{tabular}                    \\\hline 
8 & Multiple checkpoints, with fault (Eq. 6, k = 0)
& \begin{tabular}[c]{@{}l@{}}10.77\end{tabular}   
& \begin{tabular}[c]{@{}l@{}}9.50\end{tabular}   
& \begin{tabular}[c]{@{}l@{}}11.66\end{tabular}   

                          \\\hline 
9 & Multiple checkpoints, with fault (Eq. 6, k = 1)
& \begin{tabular}[c]{@{}l@{}}12.27\end{tabular}
& \begin{tabular}[c]{@{}l@{}}11.01\end{tabular}  
& \begin{tabular}[c]{@{}l@{}}13.17\end{tabular}  
\\\hline
10 & Multiple checkpoints, with fault (Eq. 6, k = 4)
& \begin{tabular}[c]{@{}l@{}}22.79\end{tabular}
& \begin{tabular}[c]{@{}l@{}}21.53\end{tabular}  
& \begin{tabular}[c]{@{}l@{}}23.67\end{tabular}  
\\\hline \hline 
11& Single checkpoint, without fault (Eq. 7)
& 10.37 & 8.99 & 11.16     \\\hline
12 & Single checkpoint, with fault (Eq. 8) & 10.87   & 9.50   & 11.66  \\\hline 
\end{tabular}
\end{adjustbox}
}
\label{tab:temp-ev-results}
\end{table}

The information {\color{black}shown} in Table~\ref{tab:temp-ev-results} allows {\color{black}us to survey} interesting aspects of SEDAR's behavior. 
As a remarkable fact, when a fault occurs, the detection mechanism (rows 4, 5 and 6) performs better than the baseline (row 2) for all the applications, regardless of the time of detection, due to the low temporal overhead implied. {\color{black}T}he sooner the error is detected, the better the mechanism behaves, as expected, because less work must be remade after stopping.
It is rather obvious that adding the multiple-checkpoint-based recovery strategy (row 7) involves a larger overhead if compared to the {\color{black}detection-only} strategy, in absence of faults; the time spent in checkpointing is worth more in the case of these long-lasting executions, but may have a not-negligible impact in shorter programs.

The analysis of the values in rows 8, 9 and 10 reveals that, when an error takes place, even rolling back several times is advantageous {\color{black}in} respect to the baseline; {\color{black}as long as} the number of rollbacks is greater than 4, the time spent in reworking is longer than the baseline strategy. 

The values shown in rows 11 and 12 are similar to the ones of rows 7 and 8{\color{black}. A}s expected, the time of recovery from the last valid application-level checkpoint (Equation~\ref{eqn:tfp4}) is almost equal to the time of recovery from the last system-level checkpoint, when {\color{black}this} is possible (Equation~\ref{eqn:tfp3} with $k$ = 0).

{\color{black}In} conclusion, it can be seen that the temporal behavior of each SEDAR strategy is dependent on the communication pattern, the computation-to-communication ratio and the detection latency. The examples described, with the three benchmarks, demonstrate how the model can be applied for temporal evaluation if the involved parameters are available (or they can be measured).

Another remarkable fact that can be observed from Table~\ref{tab:temp-ev-results} is that the different alternatives {\color{black}of SEDAR} offer considerable gains both in time and reliability, when facing the occurrence of a silent error. This item becomes particularly important in executions that can last many hours. {\color{black}Moreover}, the longer the execution time ($T_{prog}$), the more useful the fault-tolerance strategy is, because the failures are more likely to happen. As {\color{black}previously stated}, the protection mechanism should be used used for long programs: if the execution is too short, checkpoints become {\color{black}worthless}.
Despite {\color{black}the fact that} these examples cannot be taken as general conclusions, they are illustrative {\color{black} of} the potential of SEDAR in helping users of scientific applications to reach reliable executions, as they are representative of the scientific parallel applications.


\subsection{Convenience of Saving Multiple Checkpoints for Recovery}
\label{subsection:convenience}

As {\color{black}mentioned before}, in a system that saves a chain of checkpoints for rollback, the recovery is then possible after one or more attempts. However, {\color{black}due to the checkpointing and rolling-back overhead, there are possible scenarios in which} the time spent in those attempts could be longer than simply stopping upon detection and relaunching from the beginning. Therefore, it is useful to evaluate {\color{black} the convenience of saving multiple checkpoints{\color{black}. Moreover}, the benefit{\color{black}s} of using the checkpoint-based protection should be {\color{black}considered}.} 

This study is suggestive about how to use the developed model. If statistics about the frequency and typical behavior of the faults are available for a particular system that runs an application (i.e. when the faults are more likely to appear), the strategy of protection can be properly tuned between only detection and checkpoint{\color{black}ing} for recovery. 

For different values or parameter $X$, some quantitative knowledge can be extracted from the evaluation of the execution times in Equation~\ref{eqn:tfp2} and in Equation~\ref{eqn:tfp3}. 
It can be shown that the fourth term in Equation~\ref{eqn:tfp3} is equivalent to: 

\begin{equation} (\sum_{m=0}^{k}(k-m+1/2))t_i = \frac{(k + 1)^{2}}{2}t_i \end{equation}
so that Equation~\ref{eqn:tfp3} can be rewritten as:
\begin{equation}
\label{eqn:tfp5}
\begin{split}
T_{FP}=T_{prog}(1+f_d) + T_{comp} + (n+k)t_{cs} +\\+ \frac{(k + 1)^{2}}{2}t_i+(k+1)T_{rest}
\end{split}
\end{equation}

To illustrate this idea, we have selected Jacobi's method as the test case (although the procedure can easily {\color{black}be} applied to the other benchmarks){\color{black}. For this}, we have evaluated the Equation~\ref{eqn:tfp2} with three different values of $X$: if the fault is detected near the beginning ($X$ = 30\%), {\color{black}in} the middle ($X$ = 50\%) or toward the end ($X$ = 80\%) of the execution. The times obtained are compared with the ones derived from Equation~\ref{eqn:tfp5}, taking into account the following considerations.

 The reference time for this analysis is the one from Equation~\ref{eqn:tfa2}{\color{black}:} the duration of an execution using the detection mechanism, without faults. In such a total execution time (8.97 hs, see Table~\ref{tab:temp-ev-results}), $X$ = 30\% means that the fault is detected at $t$ = 2.69 hs. At this time, with $t_i$ = 1 hour, only the two first checkpoints (CK0 and CK1) have been stored. Therefore, recovery must be possible, either from CK1 (i.e. $k$ = 0) or from CK0 (i.e. $k$ = 1), so both values are admissible in Equation~\ref{eqn:tfp5}. {\color{black} The same reasoning has been followed for the other values of $X$.}

Table~\ref{tab:exec-times-X} summarizes this data. Based on the value of parameter $X$, the second column shows the times obtained with detection, safe-stop and relaunching from the start (i.e. from Equation~\ref{eqn:tfp2}), the third column shows the times obtained by rewinding to the last {\color{black}(by the moment) stored} checkpoint (i.e. from Equation~\ref{eqn:tfp5} with $k$ = 0), the fourth column shows the times obtained by rewinding to the last but one checkpoint (i.e. from Equation~\ref{eqn:tfp5} with $k$ = 1), and so on. The value $NA$ means that the current value of $k$ is not admissible in Equation~\ref{eqn:tfp5} for the current value of $X$, because the corresponding checkpoint has not been stored yet by that moment of the progress of the execution.

\begin{table}[]
\caption{Execution time with the fault detected at $X$, with only detection and with different number of rollback attempts ($k$+1)}
\begin{tabular}{|c|c|c|c|c|c|c|}
\hline
\multirow{2}{*}{$X$ {[} \%{]}} & \multirow{2}{*}{Only detection {[}hs{]}} & \multicolumn{5}{c|}{$k+1$ rollback attempts {[}hs{]}}                                                  \\ \cline{3-7} 
                             &                                          & k=0                  & k=1                    & k=2                    & k=3                    & k=4   \\ \hline
30                           & 11.66                                    & \multirow{3}{*}{9.5} & \multirow{3}{*}{11.01} & \multicolumn{3}{c|}{$NA$ (Not Admissible)}                 \\ \cline{1-2} \cline{5-7} 
50                           & 13.46                                    &                      &                        & \multirow{2}{*}{13.52} & \multirow{2}{*}{17.02} & $NA$    \\ \cline{1-2} \cline{7-7} 
80                           & 16.16                                    &                      &                        &                        &                        & 21.53 \\ \hline
\end{tabular}
\centering
\label{tab:exec-times-X}
\end{table}

For the analyzed values of $X$, the obtained results suggest that rolling back to the last stored checkpoint ($k$ = 0), if possible, is always {\color{black}advisable}, faced {\color{black}with} the case of stopping, notifying the error and relaunching from the beginning. Even to restart from {\color{black}the} last but one checkpoint ($k$ = 1) is still convenient (obviously, if that checkpoint is not corrupted). However, if the fault is detected {\color{black}around} the middle of execution, and two (or more) rollbacks have to made, it would have been preferable to stop and relaunch. In other words, if recovery from the last two checkpoints is not possible, trying from the previous ones is still more expensive than simply halting and getting started again. This is caused by the large overheads involved in not only re-executing the same computation several times, but also re-storing checkpoints and making various restart attempts.
This trend also holds as the application progresses: if the fault is detected close to the end, {\color{black}even trying more rollbacks to recorded checkpoints could represent an improvement compared to stop and relaunch.

Of course, there is no way {\color{black}of knowing} which checkpoint would enable recovery. However, following this line of reasoning,} if we force the time of Equation~\ref{eqn:tfp2} to be minor or equal to the one of Equation~\ref{eqn:tfp5} with $k$ = 0, we obtain {\color{black}$X \leq\ 5.88\%$}. This means that, before that level of progress {\color{black} (which represents about 32 minutes with this set of parameters){\color{black},} it is not convenient to record any checkpoint; it is less expensive simply {\color{black}to} stop and relaunch.} On the other hand, if we force the time of Equation~\ref{eqn:tfp2} to be greater or equal to the one of Equation~\ref{eqn:tfp5} with $k$ = 1, we obtain $X \geq\ 22.67\%$, which represents about 2 hours{\color{black}. T}his means that only when that percentage of execution time has elapsed, rolling back to the last but one checkpoint is preferable than stopping and relaunching. Before that, it is worth {\color{black}storing} only a single checkpoint. This result reinforces the idea {\color{black}that} this strategy is not so useful if the overall execution time is too short: the time required for storing the checkpoint could be {\color{black}non}-negligible if compared with the required checkpoint interval. As a final example, when $X \geq\ 50.61\%$, rolling back up two checkpoints is beneficial {\color{black}when} compared to us{\color{black}ing} the detection{\color{black}-only} mechanism. 

Despite {\color{black}the fact that this} cannot be taken as a general conclusion, the analysis above shows that (with these parameters), the overhead associated with rolling back and re-executing is much more significant than the low cost of saving checkpoints. This suggests that decreasing the checkpoint interval $t_i$ can be convenient, as the advantage of rolling back a shorter span exceeds the checkpointing overhead.

Once again, although these are simple examples, they are illustrative {\color{black}of} how useful conclusions can be drawn from the model of temporal behavior, which allow {\color{black}us to adapt} the protection strategy based on the knowledge of the {\color{black}system} parameters.



\section{Conclusions and Future Work}
\label{sec:conclusions}
Exascale computing presents several challenges to future generation computer systems and guaranteeing reliability is one of them.
The protection of the MPI applications at message level is a feasible and effective method for detecting, secluding and avoiding the propagation of data corruption, taking into account the deep effects that a single transient fault can cause on all processes that communicate.
In this article, SEDAR {\color{black}has been} presented {\color{black}as} a methodology for detecting and recovering from all silent errors, in an agnostic manner to the algorithms. SEDAR consists {\color{black}of} three complementary alternatives for only detection, recovery based on multiple system-level checkpoints and recovery based on a single user-level checkpoint.
{\color{black}The most remarkable conclusions are}:

\begin{itemize}
\item The functional behavior in the presence of faults {\color{black}can be analytically} described. {\color{black}We have built a model that considers all the fault scenarios on a well-known test application and SEDAR's response facing each scenario, thus} showing the validity of the detection and the recovery mechanisms. 
\item The predictions of the model {\color{black}can be} empirically verified.
{\color{black}Through} controlled fault injection experiments, the reliability provided by SEDAR strategies {\color{black}has been demonstrated}.
\item {\color{black} The temporal behavior of each SEDAR strategy can be characterized. When o}btaining the execution parameters {\color{black}for} applications with different communication patterns and computation-to-communication ratios{\color{black}, it has been} shown that the different {\color{black}variants} of SEDAR offer benefits both in execution time and reliability{\color{black}. This} becomes particularly profitable in {\color{black} long-lasting programs}.
\item SEDAR {\color{black}can} be adapted to a determined cost-performance trade-off. As each SEDAR strategy supplies a particular coverage but also has limitations and implementation costs, choosing between them allows {\color{black}us to adjust} to the needs of a particular system.
\item {\color{black} The temporal characterization can be used to extract useful protection guidelines. To illustrate this, it has been shown when it is beneficial to employ each SEDAR strategy}. 
\item {\color{black}Both the viability and efficacy to tolerate transient faults in expected HPC exascale systems have been shown}.

\end{itemize}

As ongoing and future lines of work, we can enumerate:

\begin{itemize}

\item {\color{black} Emulating non-deterministic calls, which are required to extend the scope of applications that can be protected with SEDAR}. 
\item {\color{black} Performing experimental validation with customized, non-coordinated user-level checkpoints, calculating the optimal checkpoint interval to minimize execution overhead, and measuring the relationship between the latency of detection and the communication pattern}. 
\item {\color{black} Refining the multiple checkpoint-based recovery mechanism to optimally support various faults, and analytically modeling the temporal response in the presence of multiple non-related faults}.
\item {\color{black}Implementing an automatic adaptation of the recovery strategy, i.e. dynamically starting protection depending on the progress of the execution (based on the reasoning stated in section \ref{subsection:convenience})}.
\end{itemize}

As a final aim, integration with scalable architectures that use C/R for permanent fault tolerance \cite{castro2015fault} should be {\color{black}attempted}. As SEDAR also provides scalable options for detection and recovery, fault-tolerance for both types of errors could be achieved for projected exascale systems. It is important to clarify that a production version of SEDAR is being developed, whereas the current implementation remains as a prototype. 

\subsection*{Acknowledgments} 
This research has been supported by the Agencia Estatal de Investigación (AEI), Spain and the Fondo Europeo de Desarrollo Regional (FEDER) UE, under contract TIN2017-84875-P and partially funded by a research collaboration agreement with the Fundacion Escuelas Universitarias Gimbernat (EUG). In addition, this research has been supported by the Universidad Nacional de La Plata, Argentina, through the Programas de Incentivos. We also would like to thank the reviewers {\color{black} and the editors} for their constructive feedback and significant contributions to this work.

\label{sect:bib}

\bibliographystyle{IEEEtran}
\bibliography{easychair}

\begin{thebibliography}{10}
\providecommand{\url}[1]{#1}
\csname url@samestyle\endcsname
\providecommand{\newblock}{\relax}
\providecommand{\bibinfo}[2]{#2}
\providecommand{\BIBentrySTDinterwordspacing}{\spaceskip=0pt\relax}
\providecommand{\BIBentryALTinterwordstretchfactor}{4}
\providecommand{\BIBentryALTinterwordspacing}{\spaceskip=\fontdimen2\font plus
\BIBentryALTinterwordstretchfactor\fontdimen3\font minus
  \fontdimen4\font\relax}
\providecommand{\BIBforeignlanguage}[2]{{%
\expandafter\ifx\csname l@#1\endcsname\relax
\typeout{** WARNING: IEEEtran.bst: No hyphenation pattern has been}%
\typeout{** loaded for the language `#1'. Using the pattern for}%
\typeout{** the default language instead.}%
\else
\language=\csname l@#1\endcsname
\fi
#2}}
\providecommand{\BIBdecl}{\relax}
\BIBdecl

\bibitem{martsin2015}
T.~Martsinkevich, O.~Subasi, O.~Unsal, F.~Cappello, and J.~Labarta,
  ``Fault-tolerant protocol for hybrid task-parallel message-passing
  applications,'' in \emph{Cluster Computing (CLUSTER), 2015 IEEE International
  Conference on}.\hskip 1em plus 0.5em minus 0.4em\relax IEEE, 2015, pp.
  563--570.

\bibitem{benoit2019combining}
A.~Benoit, A.~Cavelan, F.~M. Ciorba, V.~Le~F{\`e}vre, and Y.~Robert,
  ``Combining checkpointing and replication for reliable execution of linear
  workflows with fail-stop and silent errors,'' \emph{International Journal of
  Networking and Computing}, vol.~9, no.~1, pp. 2--27, 2019.

\bibitem{elliott2014}
J.~Elliott, M.~Hoemmen, and F.~Mueller, ``Evaluating the impact of sdc on the
  gmres iterative solver,'' in \emph{Parallel and Distributed Processing
  Symposium, 2014 IEEE 28th International}.\hskip 1em plus 0.5em minus
  0.4em\relax IEEE, 2014, pp. 1193--1202.

\bibitem{benoit2018coping}
A.~Benoit, A.~Cavelan, F.~Cappello, P.~Raghavan, Y.~Robert, and H.~Sun,
  ``Coping with silent and fail-stop errors at scale by combining replication
  and checkpointing,'' \emph{Journal of Parallel and Distributed Computing},
  vol. 122, pp. 209--225, 2018.

\bibitem{fiala2012}
D.~Fiala, F.~Mueller, C.~Engelmann, R.~Riesen, K.~Ferreira, and R.~Brightwell,
  ``Detection and correction of silent data corruption for large-scale
  high-performance computing,'' in \emph{Proceedings of the International
  Conference on High Performance Computing, Networking, Storage and
  Analysis}.\hskip 1em plus 0.5em minus 0.4em\relax IEEE Computer Society
  Press, 2012, p.~78.

\bibitem{shye2009}
A.~Shye, J.~Blomstedt, T.~Moseley, V.~J. Reddi, and D.~A. Connors, ``{PLR}: A
  software approach to transient fault tolerance for multicore architectures,''
  \emph{IEEE Transactions on Dependable and Secure Computing}, vol.~6, no.~2,
  pp. 135--148, 2009.

\bibitem{cappello2014}
F.~Cappello, A.~Geist, W.~Gropp, S.~Kale, B.~Kramer, and M.~Snir, ``Toward
  exascale resilience: 2014 update,'' \emph{Supercomputing frontiers and
  innovations}, vol.~1, no.~1, pp. 5--28, 2014.

\bibitem{lu2013}
G.~Lu, Z.~Zheng, and A.~A. Chien, ``When is multi-version checkpointing
  needed?'' in \emph{Proceedings of the 3rd Workshop on Fault-tolerance for HPC
  at extreme scale}.\hskip 1em plus 0.5em minus 0.4em\relax ACM, 2013, pp.
  49--56.

\bibitem{mushtaq2013}
H.~Mushtaq, Z.~Al-Ars, and K.~Bertels, ``Efficient software-based fault
  tolerance approach on multicore platforms,'' in \emph{Proceedings of the
  Conference on Design, Automation and Test in Europe}.\hskip 1em plus 0.5em
  minus 0.4em\relax EDA Consortium, 2013, pp. 921--926.

\bibitem{bosilca2009algorithm}
G.~Bosilca, R.~Delmas, J.~Dongarra, and J.~Langou, ``Algorithm-based fault
  tolerance applied to high performance computing,'' \emph{Journal of Parallel
  and Distributed Computing}, vol.~69, no.~4, pp. 410--416, 2009.

\bibitem{montezanti2017methodology}
D.~Montezanti, A.~De~Giusti, M.~Naiouf, J.~Villamayor, D.~Rexachs, and
  E.~Luque, ``A methodology for soft errors detection and automatic recovery,''
  in \emph{2017 International Conference on High Performance Computing \&
  Simulation (HPCS)}.\hskip 1em plus 0.5em minus 0.4em\relax IEEE, 2017, pp.
  434--441.

\bibitem{mukherjee2005soft}
S.~S. Mukherjee, J.~Emer, and S.~K. Reinhardt, ``The soft error problem: An
  architectural perspective,'' in \emph{High-Performance Computer Architecture,
  2005. HPCA-11. 11th International Symposium on}.\hskip 1em plus 0.5em minus
  0.4em\relax IEEE, 2005, pp. 243--247.

\bibitem{monte2014}
D.~Montezanti, E.~Rucci, D.~Rexachs, E.~Luque, M.~Naiouf, and A.~De~Giusti, ``A
  tool for detecting transient faults in execution of parallel scientific
  applications on multicore clusters,'' \emph{Journal of Computer Science \&
  Technology}, vol.~14, pp. 32--38, 2014.

\bibitem{chen2011}
Z.~Chen, ``Algorithm-based recovery for iterative methods without
  checkpointing,'' in \emph{Proceedings of the 20th international symposium on
  High performance distributed computing}.\hskip 1em plus 0.5em minus
  0.4em\relax ACM, 2011, pp. 73--84.

\bibitem{shantharam2012fault}
M.~Shantharam, S.~Srinivasmurthy, and P.~Raghavan, ``Fault tolerant
  preconditioned conjugate gradient for sparse linear system solution,'' in
  \emph{Proceedings of the 26th ACM international conference on
  Supercomputing}.\hskip 1em plus 0.5em minus 0.4em\relax ACM, 2012, pp.
  69--78.

\bibitem{engel2011}
C.~Engelmann and S.~B{\"o}hm, ``Redundant execution of hpc applications with
  {MR-MPI},'' in \emph{Proceedings of the 10th IASTED International Conference
  on Parallel and Distributed Computing and Networks (PDCN)}, 2011, pp. 15--17.

\bibitem{ferreira2011}
K.~Ferreira, R.~Riesen, R.~Oldfield, J.~Stearley, J.~Laros, K.~Pedretti, and
  T.~Brightwell, ``r{MPI}: increasing fault resiliency in a message-passing
  environment,'' \emph{Sandia National Laboratories, Albuquerque, NM, Tech.
  Rep. SAND2011-2488}, 2011.

\bibitem{yalcin2013}
G.~Yalcin, O.~S. Unsal, and A.~Cristal, ``Fault tolerance for multi-threaded
  applications by leveraging hardware transactional memory,'' in
  \emph{Proceedings of the ACM International Conference on Computing
  Frontiers}.\hskip 1em plus 0.5em minus 0.4em\relax ACM, 2013, p.~4.

\bibitem{ni2013}
X.~Ni, E.~Meneses, N.~Jain, and L.~V. Kal{\'e}, ``{ACR}: Automatic
  checkpoint/restart for soft and hard error protection,'' in \emph{Proceedings
  of the International Conference on High Performance Computing, Networking,
  Storage and Analysis}.\hskip 1em plus 0.5em minus 0.4em\relax ACM, 2013,
  p.~7.

\bibitem{ali2011redundant}
N.~Ali, S.~Krishnamoorthy, N.~Govind, and B.~Palmer, ``A redundant
  communication approach to scalable fault tolerance in pgas programming
  models,'' in \emph{Parallel, Distributed and Network-Based Processing (PDP),
  2011 19th Euromicro International Conference on}.\hskip 1em plus 0.5em minus
  0.4em\relax IEEE, 2011, pp. 24--31.

\bibitem{benoit2019replication}
A.~Benoit, T.~H{\'e}rault, V.~L. F{\`e}vre, and Y.~Robert, ``Replication is
  more efficient than you think,'' in \emph{Proceedings of the International
  Conference for High Performance Computing, Networking, Storage and
  Analysis}.\hskip 1em plus 0.5em minus 0.4em\relax ACM, 2019, p.~89.

\bibitem{montezanti2016characterizing}
D.~M. Montezanti, D.~Rexachs~del Rosario, E.~Rucci, E.~Luque~Fad{\'o}n,
  M.~Naiouf, and A.~E. De~Giusti, ``Characterizing a detection strategy for
  transient faults in hpc,'' in \emph{Computer Science \& Technology Series.
  XXI Argentine Congress of Computer Science. Selected papers}.\hskip 1em plus
  0.5em minus 0.4em\relax Editorial de la Universidad Nacional de La Plata
  (EDULP), 2016, pp. 77--90.

\bibitem{panadero2017p3s}
J.~Panadero, A.~Wong, D.~Rexachs, and E.~Luque, ``P3s: A methodology to analyze
  and predict application scalability,'' \emph{IEEE Transactions on Parallel
  and Distributed Systems}, vol.~29, no.~3, pp. 642--658, 2017.

\bibitem{puzyrev2015review}
V.~Puzyrev and J.~M. Cela, ``A review of block krylov subspace methods for
  multisource electromagnetic modelling,'' \emph{Geophysical Journal
  International}, vol. 202, no.~2, pp. 1241--1252, 2015.

\bibitem{swift2005}
G.~A. Reis, J.~Chang, N.~Vachharajani, R.~Rangan, and D.~I. August, ``{SWIFT}:
  Software implemented fault tolerance,'' in \emph{Proceedings of the
  international symposium on Code generation and optimization}.\hskip 1em plus
  0.5em minus 0.4em\relax IEEE Computer Society, 2005, pp. 243--254.

\bibitem{cao2016system}
J.~Cao, K.~Arya, R.~Garg, S.~Matott, D.~K. Panda, H.~Subramoni, J.~Vienne, and
  G.~Cooperman, ``System-level scalable checkpoint-restart for petascale
  computing,'' in \emph{2016 IEEE 22nd International Conference on Parallel and
  Distributed Systems (ICPADS)}.\hskip 1em plus 0.5em minus 0.4em\relax IEEE,
  2016, pp. 932--941.

\bibitem{ansel2009dmtcp}
J.~Ansel, K.~Arya, and G.~Cooperman, ``{DMTCP}: Transparent checkpointing for
  cluster computations and the desktop,'' in \emph{Parallel \& Distributed
  Processing, 2009. IPDPS 2009. IEEE International Symposium on}.\hskip 1em
  plus 0.5em minus 0.4em\relax IEEE, 2009, pp. 1--12.

\bibitem{jacobiAndrews}
G.~Andrews, ``Scientific computing,'' in \emph{Foundations of Multithreaded,
  Parallel and Distributed Computing}.\hskip 1em plus 0.5em minus 0.4em\relax
  Addison-Wesley, 2000, ch.~11, pp. 527--585.

\bibitem{rucci2011pdpta}
\BIBentryALTinterwordspacing
E.~Rucci, A.~De~Giusti, and F.~Chichizola, ``Parallel smith-waterman algorithm
  for dna sequences comparison on different cluster architectures,'' in
  \emph{Proceedings of the International Conference on Parallel and Distributed
  Processing Techniques and Applications (PDPTA'11)}, vol.~1.\hskip 1em plus
  0.5em minus 0.4em\relax WorldComp, 2011, pp. 666--672. [Online]. Available:
  \url{http://worldcomp-proceedings.com/proc/p2011/PDP5014.pdf}
\BIBentrySTDinterwordspacing

\bibitem{fialho2011missing}
L.~Fialho, D.~Rexachs, and E.~Luque, ``What is missing in current checkpoint
  interval models?'' in \emph{2011 31st International Conference on Distributed
  Computing Systems}.\hskip 1em plus 0.5em minus 0.4em\relax IEEE, 2011, pp.
  322--332.

\bibitem{daly2006higher}
J.~T. Daly, ``A higher order estimate of the optimum checkpoint interval for
  restart dumps,'' \emph{Future generation computer systems}, vol.~22, no.~3,
  pp. 303--312, 2006.

\bibitem{castro2015fault}
M.~Castro-Le{\'o}n, H.~Meyer, D.~Rexachs, and E.~Luque, ``Fault tolerance at
  system level based on {RADIC} architecture,'' \emph{Journal of Parallel and
  Distributed Computing}, vol.~86, pp. 98--111, 2015.

\end{thebibliography}

\end{document}